# The Birth of Radar Meteor Astronomy at Jodrell Bank: The Collaboration between Bernard Lovell and Manning Prentice

**Jeremy Shears**

## Abstract

The discovery of radar echoes from meteor trains at Jodrell Bank in December 1945 heralded the beginning of a new era in meteor astronomy. This paper examines the early collaboration between Bernard Lovell, whose expertise in radar and physics transformed wartime technology into a new astronomical technique, and J. P. Manning Prentice, Director of the British Astronomical Association Meteor Section, whose knowledge of visual meteor observation proved essential to interpreting the discoveries. Drawing on previously unpublished correspondence from the Lovell Archive at the University of Manchester, the paper traces the development of their partnership, including the pioneering 1946 Perseid and Giacobinid observations, the contribution of BAA observers, and the establishment of radar meteor research at Jodrell Bank. It reveals how complementary professional and amateur expertise combined to create a new field of astronomy and a lasting scientific friendship.

## Introduction

The establishment of the radar meteor programme at the University of Manchester in the years immediately following the Second World War marked the birth of a new branch of astronomy. Led by Bernard Lovell (Alfred Charles Bernard Lovell, 1913-2012), the programme demonstrated for the first time that radar could be used to detect and study meteors day and night, irrespective of cloud, transforming the observation of meteors from an activity dependent upon visual observers into one capable of generating thousands of objective measurements. These pioneering investigations laid the foundations not only for radar meteor astronomy, but also for the wider development of radio astronomy at Jodrell Bank.

Histories of Jodrell Bank have understandably concentrated on Lovell's scientific vision and his pioneering use of wartime radar equipment for astronomical research. Much less attention has been paid to the contribution made by Britain's amateur astronomers, whose expertise in meteor observation proved essential during the formative years of the programme. Among these, one individual stands out: John Philip Manning Prentice (1903-1981), Director of the British Astronomical Association Meteor Section from 1923 to 1954.

Drawing upon previously unpublished correspondence between Bernard Lovell and Manning Prentice, held in the Lovell Archive at the John Rylands Library of the University of Manchester, which contains over 50 letters and documents dated been 2 June 1946 and 18 December 1950, this paper examines the origins of radar meteor astronomy at Jodrell Bank. The correspondence reveals that Prentice's role extended far beyond providing visual observations. He introduced Lovell to many of the principal scientific questions then confronting meteor astronomy, organised the network of amateur observers that validated the early radar experiments, contributed to the interpretation of radar observations and became one of Lovell's closest scientific collaborators during the formative years of the Jodrell Bank programme.

The paper argues that the success of the early radar meteor programme depended upon a remarkable partnership between professional physicists and experienced amateur astronomers. Radar supplied an entirely new observational technique, but its scientific value could only be realised through comparison with visual observations accumulated over decades by the amateur community. Manning Prentice became the crucial bridge between these two traditions, helping to shape both the scientific direction of the Jodrell Bank programme and the emergence of radar meteor astronomy as a new discipline.

By examining the evolution of the Lovell–Prentice collaboration, this paper offers a fresh perspective on the early history of Jodrell Bank. It demonstrates that the birth of radar meteor astronomy was not simply a technological achievement arising from wartime radar, but a collaborative enterprise in which amateur and professional astronomers made complementary—and equally essential—contributions.

**The origins of Bernard Lovell's Jodrell Bank vision**

Bernard Lovell was born in a village on the outskirts of Bristol. From an early age he developed a fascination with radio, building crystal receivers and his own transmitter while still at school. Although an able student, he found school uninspiring until, at the age of fifteen, he attended a public lecture at the University of Bristol featuring spectacular demonstrations of electrical discharges, by Professor A. M. Tyndall. The experience inspired him to pursue physics, and from then on his academic performance improved dramatically.

Lovell studied physics at the University of Bristol, where his enthusiasm for experimental science led him to assist with research during university vacations. He remained at Bristol for doctoral research under Tyndall, investigating the electrical conductivity of thin films, before seeking an academic appointment on completing his PhD.

Tyndall encouraged Lovell to broaden his experience and arranged interviews at both Birkbeck College, London, with Professor Patrick Blackett, and the University of Manchester with Professor Lawrence Bragg. Although Lovell hoped to work with Blackett on cosmic rays, only Manchester offered him a position. Initially disappointed by the move, his fortunes changed in 1937 when Blackett himself succeeded Bragg as head of the Manchester physics department. Blackett immediately invited Lovell to join his cosmic ray research group, beginning a scientific partnership that would shape the remainder of his career.

By the summer of 1939 Lovell was preparing an expedition to the Pic du Midi Observatory to study cosmic rays when international events intervened. Following

Blackett's instructions, the expedition was abandoned and Lovell was recruited into Britain's highly secret wartime radar programme.

During the Second World War Lovell spent six years developing airborne radar systems, leading a team responsible for increasingly sophisticated equipment used in aircraft navigation and anti-submarine warfare (Figure 1). The work gave him unrivalled practical experience of radar, radio engineering and electronic instrumentation. On the day war was declared, while working at RAF Staxton Wold, he also witnessed unusual echoes on a radar display that were explained as ionospheric effects. The incident remained in his memory and foreshadowed his later interest in unexpected radar phenomena.

By 1945, Lovell had become one of Britain's leading radar physicists. The knowledge and technical expertise acquired during the war would prove fundamental to the remarkable discoveries that followed at Jodrell Bank.

**Post-war: cosmic rays and radar**

At the end of the Second World War, Lovell returned to Manchester and resumed his cosmic ray research. Blackett reminded him of their wartime idea of using radar to detect cosmic ray showers, and Lovell obtained a surplus radar unit from his wartime colleague J. S. Hey.

The equipment was installed beside the physics department in central Manchester, but the experiment immediately encountered a problem: electrical interference from the trams running along Oxford Road overwhelmed the sensitive receivers. A quieter location was needed.

Professor F. W. Sansome of the Botany Department suggested that Lovell visit the University's experimental station at Jodrell Bank, about twenty miles south of Manchester. The site appeared ideal: it was free from electrical interference and isolated from the city. There were two huts on the site, used by the gardeners which would prove useful for tea breaks and warming hands over a brazier. These huts still survive today and are Grade II listed (Figure 2). Lovell obtained permission to install his equipment there for two weeks — but he never left.

Using his wartime contacts, Lovell arranged for the radar equipment to be transported to Jodrell Bank on 12 December 1945 (Figures 3 and 4). After overcoming problems with the generator, he switched on the system late on 14 December. Almost immediately, he detected strong echoes and was convinced that he had begun to observe the universe with a new technique.

The signals were not, however, caused by cosmic rays. By chance, the experiment coincided with the peak of the Geminid meteor shower, and the echoes were produced by ionised meteor trains. Lovell later recalled:

"It was dark and late, but I was elated and excited. Six years of longing and several months of doubt and frustration were behind me. Again I was a real scientist able to study the universe with a new technique." (1)

**A crisis with Blackett**

Lovell initially expected the radar to provide data on cosmic ray showers, but in early 1946 Blackett confronted him with an unexpected problem. Thomas Eckersley of the Marconi Wireless Telegraph Company had identified a serious flaw in their 1940 paper, Radio echoes and cosmic ray showers (2), which had suggested that radio reflections might be caused by cosmic ray induced ionisation.

Eckersley pointed out that the paper contained a mathematical error and, more importantly, questioned whether Lovell had considered the "damping factor". Lovell had not, and he discovered that the Manchester equipment was far too small to detect the predicted cosmic ray echoes.

Lovell later reflected that wartime pressures had contributed to the oversight: "The day I posted the draft of that paper to Blackett, we were being machine-gunned by German aircraft… and it was no time to consider damping factors high in the atmosphere." (1)

Despite this setback, Lovell and Blackett's relationship survived (Figure 5). More importantly, the unexpected meteor echoes at Jodrell Bank opened an entirely new field of research and set Lovell on the path towards radio astronomy.

**Finding the meteor expert: Herlofson introduces Manning Prentice**

Lovell continued investigating the possibility that meteor trains were responsible for the mysterious radar echoes detected at Jodrell Bank. Neither he nor Blackett had any detailed knowledge of meteor science, leading Blackett to suggest that they consult Nicolai Herlofson (1916–2004, Figure 6).

"After all," Blackett remarked, "Herlofson is a meteorologist — he's bound to know about meteors!"

Lovell was unsure whether Blackett was joking. In fact, Herlofson was a physicist at Manchester who, after escaping Norway during the war, had worked as a meteorologist. He later became internationally recognised for his work in astrophysics, including, with Hannes Alfvén, the prediction of synchrotron radiation from the Milky Way.

Herlofson knew that much of the leading work on meteors was being carried out not by professional astronomers but by skilled amateurs. He therefore contacted the British Astronomical Association (BAA), whose Secretary, C. F. N. Powell, introduced him to J. P. Manning Prentice (Figure 7).

Prentice, a solicitor from Stowmarket, had joined the BAA in 1919 and became Director of the Meteor Section in 1923 at the age of twenty. By 1946 he was widely regarded as the leading authority on visual meteor observation in Britain. His dedication was legendary: during a brief break from an all-night meteor observing session in 1934, he discovered Nova Herculis, now known as DQ Herculis.

There was arguably no one in Britain better placed to advise Lovell on meteors. Herlofson's first telephone conversation with Prentice took place on 2 June 1946. He kept detailed notes of these early discussions, which he used to brief both Lovell and Blackett. Among Prentice's key points was that reliable meteor observations required extensive training: "6–12 months training under good supervision is necessary for an observer before he can take observations useful for height determination" (3).

Prentice also described the network of experienced observers he had developed over many years, mentioning George Alcock (1912–2000), Neville Goodman (1917–2001), Harold Ridley (1919–1995) and R. Manners (4).

From the outset, Prentice recognised the potential of radar meteor observations and was eager to support the work at Jodrell Bank. He proposed that BAA Meteor Section observers should travel to Jodrell for the forthcoming Perseid meteor shower in 1946. By the end of June he had prepared an observing schedule including Alcock, Ridley, M. W. Ovenden, G. S. Hawkins and J. D. P. Williams (5) (Figure 8).

Herlofson relayed Lovell's concern that technical difficulties might prevent the radar system being ready in time, rendering the observers' journey unnecessary. Prentice nevertheless argued that the opportunity was too important to miss. His observers had limited availability because of employment or study commitments, but he assured Lovell that "BAA observers fully understand the experimental and uncertain nature of the work" (6).

On 14 July 1946, Blackett approved travel expenses for the BAA observers, authorising up to £60 from his research budget (7). The letter, written while Blackett was travelling to a conference in Cambridge, represents an important milestone in the development of the Jodrell Bank meteor programme (Figure 9): (7)

> "Dear Herlofson
>
> Yes please go ahead with the meteor observation plans. Will you tell Lovell. I note that the cost is likely to be up to £60. This can come out of the Nuffield Fund..."

Herlofson's contribution to the emergence of Lovell's meteor programme deserves greater recognition. Without his intervention, the connection with Prentice — which became central to the success of the early Jodrell Bank observations — might never have been made.

Herlofson also remained actively involved as the programme developed and contributed to the early scientific interpretation of meteor radar observations. In 1947 he was a co-author, alongside Prentice, Lovell, R. W. B. Pearse and J. G. Porter, of a major review of meteor science published from Jodrell Bank (8). Each author contributed a section: Herlofson discussed the theory of meteor ablation and train ionisation, while Prentice introduced the history of meteor observation and summarised the state of knowledge before the new radar techniques emerged.

**Successful observations of the 1946 Perseids**

In his autobiography, *Astronomer by Chance* (1), Lovell recalls Prentice arriving at Jodrell Bank for the 1946 Perseids in an open-top car, the back seat laden with a deckchair, flying suit, flashlights, a large celestial globe and various charts. He gave a similar account in his later oral history interviews (9).

The contemporary correspondence, however, tells a different story. It shows that Prentice was not present at Jodrell Bank during the 1946 Perseids. Instead, he was on holiday in Southwold, on the Suffolk coast, during the period of maximum activity, honouring a long-standing commitment to accompany his Boys' Brigade contingent. His choice of dates appears to have been deliberate. Full Moon occurred on 12 August and, like many experienced meteor observers, Prentice generally avoided organising observing campaigns around full Moon because of the unfavourable observing conditions. The Perseid maximum therefore provided a convenient opportunity to take his holiday.

The correspondence also clarifies which members of the BAA Meteor Section participated in the observing campaign at Jodrell Bank. The paper describing the 1946 Perseid observations in *Monthly Notices of the Royal Astronomical Society* (10) credits two young observers, Gerald Hawkins (1928– 2003) and Michael Ovenden (1926–1987) who were 18 and 20 years old respectively. This is entirely consistent with the archival evidence. By contrast, George Alcock, Harold Ridley and J. D. P. Williams, all of whom appeared in Prentice's original observing schedule, did not attend.

Lovell wrote to Prentice on 12 August, following the night of Perseid maximum, to update him on the radar and visual observations of the Perseids and followed this up with a telephone call. It had been a great success, prompting Prentice to write a few days later:

“There is clearly no doubt that your echoes lasting over 0.25 second are connected with visual meteors….” (11)

Prentice went on to suggest that shorter echoes might well be due to meteors below the limit of visual detection.

The correspondence also establishes when Prentice first visited Jodrell Bank. Contrary to Lovell's later recollections, his first visit did not take place until the weekend of 7–8 September 1946. It was on this occasion that Prentice and Lovell met in person for the first time.

Their discussions that weekend centred largely on preparing a paper describing the 1946 Perseids. Given Prentice's extensive experience in meteor astronomy, it was natural that he should take the lead in planning the paper and developing its structure.

On returning home, Prentice wrote to thank Lovell for his hospitality and enclosed an outline for the paper. Lovell responded enthusiastically:

“I think your proposed outline is excellent. The next stage would be for me to draft out all our part of the paper.... After your visit .....my enthusiasm has increased even more.” (12)

Over the following weeks, Prentice and Lovell worked closely on the manuscript, exchanging drafts, suggestions and critical comments. The result, *Radio Echo Observations of Meteors* (10), was submitted to *Monthly Notices* on 8 January 1947.

Prentice was listed as the first author, followed by Lovell and C. J. Banwell, a member of the Jodrell Bank team. Their affiliations presented at the end of the paper—"Meteor Section, British Astronomical Association" and "Physical Laboratories, University of Manchester"—reflected the partnership between the amateur and professional communities that underpinned the research. The paper also acknowledged the contributions of the two BAA Meteor Section observers, Gerald Hawkins and Michael Ovenden, whose visual observations had been central to the success of the 1946 Perseid campaign.

**The Searchlight Aerial and the 1946 Giacobinids**

During 1946, Lovell accumulated substantial quantities of surplus wartime equipment, much of it collected from around the country by his newly appointed assistant, John A. Clegg (1913–1987; Figure 10).

The first major acquisition was a large ex-Army *Park Royal* vehicle packed with radar equipment. Unfortunately, while attempting to drive it across the observing field, Clegg became hopelessly bogged down, the vehicle sinking to its axles in the Cheshire clay. There it remained, serving as Lovell's first headquarters and, quite unintentionally, determining the future layout of the Jodrell Bank site. A few years later, the rusting vehicle was replaced by a permanent building that retained the name *Park Royal*, by which it is still known today.

Another important addition to the Jodrell site was the Searchlight Aerial (Figure 11), which proved crucial to the next phase of the meteor research programme. Lovell wished to direct his radar aerials towards any part of the sky during a meteor shower, both towards the radiant and at right angles to it. To achieve this, he acquired a surplus Army searchlight mounting, dispensing with the reflector but retaining its robust altazimuth mount, which was ideally suited to supporting an array of Yagi aerials.

Construction of the aerial almost came to a halt because of an unexpected shortage of timber. In the immediate post-war years, wood remained strictly rationed, making it virtually impossible to obtain a licence to purchase the quantity required for the aerial supports.

Lovell later recalled that the Jodrell gardeners "directed us to an ancient Elizabethan water mill in a neighbouring village" (1). The owner, aware of the work being undertaken at Jodrell Bank, generously supplied the necessary timber and never sent an invoice.

The mill is not identified in Lovell's account. However, the archival evidence suggests that it was almost certainly Bate Mill, at Peover Superior, approximately one kilometre from Jodrell Bank (Figure 12). Although now disused and partly derelict, it represents another small but tangible link with the earliest days of the observatory.

The Searchlight Aerial was commissioned in time for the Giacobinid meteor shower of 9 October 1946. Also known as the Draconids, Prentice referred to this as “my shower” (13). His observations of an outburst in 1926 confirmed that it was associated with comet 21P/Giacobini–Zinner. A further outburst occurred in 1933, and Prentice anticipated another meteor storm in 1946 to which he enthusiastically alerted Lovell as a great experimental opportunity to confirm that the radar echoes were associated with meteor trains.

During the storm, meteor rates briefly reached around 10,000 per hour, both visually and by radar (13). The decisive experiment took place close to the time of maximum activity. When the Searchlight Aerial was directed towards the radiant, the number of radar echoes fell dramatically. Rotating the aerial through 90° restored the echo rate, demonstrating that the radar was detecting specular reflections from ionised meteor trains rather than the meteoroids themselves. The steerable aerial also enabled the radiant of the shower to be determined accurately and followed as it moved across the sky.

Although Prentice did not travel to Jodrell Bank for the observations, choosing instead to observe from his home in Suffolk, he remained closely involved in the scientific interpretation of the results. The subsequent paper, *Radio echo observations of the Giacobinids meteors, 1946* (14), was actually communicated to *Monthly Notices* by Prentice who also advised Lovell on the analysis. The published paper acknowledged his role in recognising the scientific importance of the predicted outburst and encouraging the observations.

The Searchlight Aerial itself had been acquired only on loan from the Army, which periodically requested its return. The rapid development of high-altitude jet aircraft, however, rendered military searchlights increasingly obsolete, allowing the structure to remain at Jodrell Bank. Its base survives today (Figure 13), where it is preserved as a Grade II listed structure.

**A deepening collaboration**

On 13 December 1946, Prentice and Lovell jointly presented the results of the Giacobinid observations at a meeting of the Royal Astronomical Society in London. The presentation generated considerable discussion, with numerous questions from the audience. In summing up the meeting, the President, Professor H. H. Plaskett, observed that the speakers had "presented to us an entirely new field of astronomical research". (15)

For Lovell, the meeting represented a turning point. It marked the first public recognition of radar meteor astronomy by the wider astronomical community, and he later recalled his delight at the reception the work had received:

“At the beginning of the meeting on that Friday afternoon we were strangers to the astronomers, aliens infiltrating a privileged assembly. As we showed our slides, the mood changed.... by the end of the meeting we were part of the astronomical community.” (1)

Lovell also noticed the effect the collaboration was also having on Prentice:

“With the correlation between the radio echoes and meteors finally established, Manning Prentice’s enthusiasm was redoubled. At last he could find out what the meteors were doing on cloudy and moonlit nights, and for our part we badly needed his visual correlations to help us unravel the complexities of the echoes. So it was that Prentice became a frequent visitor.” (1)

Following the success of the Giacobinids, the collaboration between Lovell and Prentice deepened. Prentice next observed the Leonids at Jodrell Bank in November 1946 (13).

“So it was that Prentice became a frequent visitor, and that winter [of 1946/47] we understood the reason for his flying suit. A deck chair at 4 a.m. on a snow-covered field near the time of the Geminid meteors in mid-December is a severe test of a man’s observing ability,” observed Lovell. (1)

The winter of 1946–47 was one of the coldest and snowiest on record in Britain and became widely known as the "Big Freeze". From late January 1947, prolonged blizzards, temperatures as low as −21 °C in Essex and snowdrifts reaching seven metres in places brought much of the country to a standstill.

Prentice referred to the severe weather in a letter to Lovell dated 25 February 1947, written shortly after the birth of his son, Michael. He described how the doctor had been unable to reach his Suffolk home by car and had been forced to walk "two or two and a half miles" to attend the delivery (16).

Despite these difficulties, the collaboration between Prentice and Lovell continued to flourish. In March 1947 they jointly organised a course on meteor science and radar in Manchester under the auspices of the Physical Society (17). Both lectured, together with Nicolai Herlofson and J. G. Porter (1900–1981), Director of the BAA Computing Section, whose expertise in meteor trajectory calculations had become an increasingly important component of the Jodrell Bank research programme. A summary of the course, published later that year, provides an excellent overview of contemporary understanding of meteor science (8).

Over the following years the major annual meteor showers continued to be observed. Although Prentice was often unable to travel to Jodrell Bank because of the demands of his legal practice, a recurring theme in his correspondence, the radar programme continued to produce important discoveries. One morning, during observations of the η Aquariids in May 1947, the Jodrell Bank team left the radar operating after dawn had broken. To their surprise, meteor activity continued well into daylight hours (18). Lovell had inadvertently discovered the phenomenon of daytime meteor showers, opening another productive avenue of research. Further daytime streams were subsequently identified during the summer and in the years that followed (19).

Reflecting on these discoveries, Prentice later wrote: "Alas, I was unable to go to Jodrell Bank at that time: all I knew of it was a series of postcards from Professor Lovell" (13).

Prentice nevertheless remained closely involved in the programme and he “spent many happy days (and nights) at...Jodrell Bank Observatory in the course of this work” (13) . He sometimes alerted Lovell to unusual meteor activity. One example was a telegram (Figure 14) suggesting a possible Bielid outburst in November 1948. The event was successfully observed at Jodrell Bank and resulted in a paper published in the BAA *Journal* (20). Visual observations made independently by Prentice and George Alcock from their home observing sites were incorporated into the analysis, once again illustrating the continuing value of coordinated professional–amateur observations.

**The interstellar meteors controversy**

It was generally accepted among meteor scientists that shower meteors were associated with cometary debris and that the resulting meteoroids therefore followed elliptical orbits around the Sun. A more contentious debate emerged, however, during the 1940s concerning sporadic meteors. Some researchers argued that a significant proportion of these objects followed hyperbolic trajectories and therefore originated outside the Solar System. If correct, such observations would provide important information about the distribution and properties of interstellar dust.

The hyperbolic interpretation was strongly advocated by E. J. Öpik, first at Harvard and later at Armagh Observatory. In 1940, he published a comprehensive analysis of visual meteor observations in which he concluded that approximately 60% of the recorded meteors possessed hyperbolic orbits.

This interpretation was challenged by J. G. Porter of the BAA Computing Section, who argued that most apparent hyperbolic trajectories arose from observational uncertainties and mathematical errors rather than representing genuine interstellar material. Porter based his analysis on the most reliable meteor orbit data available, much of which had been obtained through the work of the BAA Meteor Section, including observations coordinated by Prentice, whose skill in determining meteor trajectories was widely recognised.

Prentice was therefore familiar with the controversy and, like Porter, was highly critical of Öpik’s conclusions. Nevertheless, he maintained an open mind and recognised that the new radar techniques being developed at Jodrell Bank might provide a means of resolving the issue. Early in 1947, he introduced Lovell to the question of interstellar meteors (21), suggesting that radar measurements could be used to determine whether high-velocity meteor streams of interstellar origin existed.

The full intensity of the controversy did not become apparent to Lovell until a meteor conference held at the University of Manchester on 7–8 September 1948. The principal figures involved were all present: Öpik, Porter, Prentice, Lovell and the distinguished Harvard astronomer, Fred L. Whipple (1906–2004).

Suddenly, the group 'stopped because a violent and bitter dispute erupted between Öpik and Porter' (1) over the interpretation of results from a photographic meteor experiment in Arizona designed to investigate possible interstellar meteors. Porter was highly critical of the experimental conclusions, a position that Öpik regarded as a serious challenge to his interpretation of the observations.

Whipple, who remained neutral in the dispute, and Lovell watched the exchange with concern. Seeking to resolve the impasse, Whipple suggested that Lovell's radar equipment might provide the necessary evidence. Precise measurements of sporadic meteor velocities could distinguish between normal Solar System meteors and genuinely interstellar objects: meteors on hyperbolic trajectories would require velocities (with respect to the Sun) exceeding 42.2 km/s (1).

Lovell accepted the challenge and began modifying the Jodrell Bank radar system to measure meteor velocities with the required accuracy. Over the following three years, an extensive programme of observations was undertaken. By its completion in December 1951, the velocities of 1,095 sporadic meteors had been measured, yet none exceeded the critical value 42.2 km/s. These results provided strong evidence against a significant population of interstellar meteors, although Öpik did not concede the argument until 1969 (22).

Porter subsequently summarised his position in his 1949 BAA Presidential Address, *The Study of Meteor Velocities*, in which he argued strongly against the existence of hyperbolic meteors based on the available observational evidence (23).

**George Alcock visits Jodrell Bank**

George Alcock was one of the most active members of the Meteor Section and had collaborated with Prentice for many years. Their observations, made from two separate sites—Alcock near Peterborough and Prentice near Stowmarket—provided some of the most accurate meteor trajectories then available through triangulation. It was therefore unsurprising that Alcock's name appeared in Prentice's first list of observers proposed to visit Jodrell Bank for the 1946 Perseids. However, as noted earlier, he did not make that visit. His name continued to appear in the correspondence as a potential observer, and it was therefore of interest to establish when he eventually travelled to Jodrell Bank.

As far as can be determined, Alcock visited only once, in September 1950. Prentice informed Lovell that Alcock would be available to come to Jodrell Bank between 5 and 10 September (24) (25), although he would need to return home near Peterborough by 11 September to resume his teaching duties. Prentice gave Lovell a characteristically warm assessment of his long-time observing colleague: “You will find him a little reserved to get on with, but he is definitely a diamond (if a little under polished) and easily the most loyal of my observing team.”

Alcock also wrote directly to Lovell before travelling to Manchester, confirming the dates of his visit and asking whether “a couple of old blankets and a deckchair” could be provided for his overnight meteor observations (26). Unfortunately, the weather proved uncooperative. According to Alcock’s biography, he found the experience disappointing: “The weather was appalling, and most of the staff there obviously considered our observing methods and equipment quite ludicrous”. (27)

Matters were not improved when, during an evening visit to a Manchester cinema, he travelled in an open jeep through heavy rain. He returned home to Peterborough early, only for the skies to clear shortly after he left.

There was, however, a positive outcome from the visit. Alcock was able to study the scientific literature in the Jodrell Bank library and review the results obtained from Meteor Section observations. Reflecting on the experience, he commented: “The BAA had published very few of our results.... it’s very hard for an observer to remain enthusiastic when he’s at work all day and observing most of the night, without any recognition of the results of his labours” (27)

Alcock’s comments reveal an important issue facing amateur meteor observers during this period: the desire not merely to collect observations, but to see those observations recognised as meaningful scientific contributions. This contrasts with the views of some other BAA members, who complained to Association officers that too much meeting time and *Journal* space was being devoted to the new meteor research at Jodrell Bank at the expense of “real astronomy” (28).

**Jodrell Bank becomes a BAA recruiting ground**

With J.G. Porter’s encouragement, Lovell was elected to the BAA on 31 December 1947, together with three other members of his team: J. A. Clegg, J. G. Davies and V. A. Hughes. Porter proposed their nominations and Prentice seconded them (29). This reflected the close links that had developed between the Jodrell Bank group and the BAA Meteor Section.

As Lovell’s research team expanded, further members also joined the BAA, including Mary Almond and Victor Hughes. In this way, the professional research group developing at Jodrell Bank became increasingly connected with the wider amateur astronomical community.

One member of Lovell's team already had strong links with the BAA. Michael Ovenden was elected on 31 March 1942. He served as Papers Secretary from 1946 to 1951. When he began the role he was only 17 making him the youngest such Secretary.

Gerald Hawkins had been an active Meteor Section observer while still at school. After studying at the University of Nottingham and University College London, he joined Jodrell Bank to undertake his PhD research. As mentioned previously, Hawkins participated in the 1946 Perseid campaign at Jodrell Bank. He was elected to the BAA on 29 January 1947.

These BAA members subsequently pursued distinguished careers in radio astronomy in their own right. Their individual achievements, however, lie beyond the scope of this paper and will be discussed elsewhere.

Hawkins' observing colleague for the 1946 Perseids, Michael Ovenden, was also a BAA member. He went on to take a PhD at Cambridge. His subsequent research career was in the study of eclipsing binary stars.

**"Dear Dr Lovell", "Dear Mr Prentice"**

The correspondence between Lovell and Prentice reveals that there was an immediate meeting of minds following Herlofson's initial introduction, which was soon followed by Prentice's first visit to Jodrell Bank in September 1946. Over the following years, their professional collaboration developed into a close personal friendship.

It is particularly revealing to see how their relationship evolved from one in which Prentice acted as Lovell's guide to meteor astronomy—providing essential knowledge, interpreting observations and taking the lead in drafting much of their first Monthly Notices paper—into a partnership of equals. This partnership was founded on mutual respect and complementary expertise: Prentice brought his unrivalled knowledge of meteors and observational astronomy, while Lovell contributed his expertise in radar, electronics and physics.

Their personal relationship also became progressively warmer. As was customary at the time, their early correspondence used formal salutations: "Dear Dr Lovell" and "Dear Mr Prentice". After about a year, these had evolved into the more familiar, "Dear Lovell" and "Dear Prentice".

After a further period of friendship, the letters began to open with "Dear Bernard" and "Dear Manning". This change appears to have been initiated by Prentice, the older man, in a letter of October 1949 (30). Lovell adopted the same tone in his reply a week later, beginning "Dear Manning" for the first time (31).

Following his first visit to Jodrell Bank, when he stayed at a local hotel, Prentice became a regular guest at the Lovell family home—first in Timperley near Manchester and later at the rural village of Swettenham, a short distance from Jodrell Bank. Indeed, Lovell's first letter (32) from his new home, "The Quinta" at Swettenham in 1948, was addressed to Prentice, and their subsequent correspondence frequently included personal family news as well as scientific matters.

Joyce Lovell, always a generous host, greatly enjoyed Prentice's visits and the opportunity they provided to exchange family news and maintain the close friendship that had developed between the two families.

**Prentice's hopes for his own meteor radar unit**

One area of increasing discussion, and occasional frustration, between Prentice and Lovell concerned access to equipment and resources.

Prentice was unable to visit Jodrell Bank as frequently as he would have liked because of his professional, church, family and other commitments. In addition, the journey from his home in Suffolk to Jodrell was a substantial undertaking, involving a distance of around 340 km. As noted earlier, he was therefore not present at Jodrell for either the Perseids or the Giacobinids in 1946.

There was another important consideration: Prentice often preferred to observe major meteor showers from his home site, particularly when the Moon phase was favourable. This also meant that he was reluctant to send his "best observers", as he described them, to Jodrell Bank. These included George Alcock, Harold Ridley and Neville Goodman, whose absence could significantly affect the results obtained by his south-eastern observing network, especially when simultaneous observations and triangulation were required. Furthermore, each observer had their own personal commitments that restricted their availability.

For these reasons, Prentice became increasingly keen to establish a radar facility closer to his home. He first raised the possibility in his initial conversation with Herlofson on 2 June 1946 (3) and later discussed the idea directly with Lovell in a letter (33).

By this stage, several BAA meteor observers had already visited Jodrell Bank, with Prentice carefully organising the arrangements and frequently accommodating last-minute changes of personnel.

With Lovell's research group expanding, Prentice suggested that a radar unit in the south-east would provide major advantages. It would allow the established BAA network of observers to work from their own locations, avoiding the difficulties of travel and making better use of simultaneous observations. He even proposed that Lovell send an expedition to Suffolk for the 1947 Perseids, bringing equipment which they could borrow from the Army.

Lovell was initially supportive of the idea of a Suffolk radar station, but practical difficulties soon became apparent, particularly regarding the availability, maintenance and funding of suitable equipment. He explained (34) that he had approached Blackett about obtaining radar equipment on loan from the Army, but Blackett considered this unlikely to succeed. Nevertheless, Blackett agreed to explore whether the Astronomer Royal might be able to provide support for purchasing equipment.

Lovell pointed out, however, that even if suitable equipment could be obtained, the challenge of finding staff able to operate and maintain it would remain.

Prentice was not easily discouraged. In his reply (35) he argued that a radar observing programme based in the south-east, operating only during selected periods, would require far less manpower than the continuous observing programme at Jodrell Bank. He even asked whether he could be trained to operate the equipment himself, expressing his willingness to learn. He was also prepared to contribute financially, offering to cover some of the equipment costs. Later, he went further and suggested that he could contribute towards the salary of a Jodrell staff member if someone could be based in Suffolk (36).

Meanwhile, Blackett was unable to secure support from the Astronomer Royal, the Royal Observatory Greenwich or the Department of Scientific and Industrial Research. Prentice had also significantly underestimated the cost of the required equipment, suggesting that "2 or 3 units" might be obtained for under £100 each (37). Lovell had to explain that the true capital cost was likely to be closer to £20,000–£30,000 (38).

Lovell attempted to manage expectations, writing: "I have no doubt that we shall get this equipment, but I do not want to make you too optimistic" (38). In the same letter, he outlined the extensive range of additional equipment that would be required.

Prentice had suitable land available at his rural home, "Star Ridge" (Figure 15), near Battisford, approximately 6 km south of Stowmarket. He envisaged installing the equipment at his house, with ample space for aerials on the surrounding land and outbuildings available for supporting facilities. He even confirmed with the local electricity board that his electrical supply would be adequate to meet the substantial power requirements of the equipment. However, Lovell realised that Prentice had underestimated the physical scale of wartime radar systems. He cautioned:

"By the way, unless you have a gigantic stairway, doorway and floor supported on girders I wouldn't try to get the receiver in your study. I am sure that your wife will never forgive you. Also these equipments have a nasty habit of giving forth smoke at midnight." (38)

Undeterred, Prentice suggested that if his own home proved unsuitable, a radar facility might instead be established at Cambridge University, which would still allow access to his network of visual observers. Lovell replied that "I do not think there is the slightest possibility for arranging for such an instrument to be erected for work under the auspices of Cambridge" (39)

The correspondence on this issue continued for several years. During this period Lovell's equipment at Jodrell Bank became increasingly sophisticated and specifically adapted to meteor research, eventually bearing little resemblance to the surplus Army radar system installed in 1945. As a result, the cost of establishing an equivalent facility elsewhere increased substantially, while the technical demands of operating it became progressively greater.

Prentice nevertheless continued to explore possible solutions. On one occasion, having heard of surplus German military radar equipment, he raised the possibility of using it at a new site. Lovell responded with a detailed explanation of why such equipment would not be suitable and, perhaps showing some frustration, emphasised the practical difficulties involved:

“it may be salutary for me to counter by asking if you appreciate the enormous difficulties in being able to operate such a complicated instrument with any regularity and at the times when one really needs it. This introduces a very big "cloud" factor in your estimates of the advantages which would accrue…” (39)

Prentice’s ambition to establish a local radar facility was never realised. Looking back, Lovell later regretted having encouraged his hopes too enthusiastically. In July 1950 he wrote: (40)

“I am very appreciative of the disappointment which you must feel….the problems and difficulties of running even one of these radio equipments is severe….In retrospect I feel that it was wrong of …. myself ever to encourage you in the belief that it would be possible to provide any sort of local radio aid under your direction in the Battisford neighbourhood.”

**Photographic detection of meteors**

Despite spending most of his observing career engaged in visual meteor work—the simplest form of meteor observation—Prentice was by no means opposed to technological innovation. On the contrary, his correspondence demonstrates considerable enthusiasm for the ways in which radar and photographic techniques were transforming meteor astronomy.

Before the Second World War, Prentice had acquired a 10-inch (25 cm) f/1 Schmidt camera designed for wide-field (15°) photography, with the aim of recording meteor trails. The outbreak of war prevented him from developing the programme and by 1946 there was still no equivalent optical capability at Jodrell Bank. Recognising its potential value, he offered the camera to Lovell on temporary loan in August 1946 and brought it with him during his first visit to Jodrell Bank the following month (11).

By 1949, Lovell’s group began experimenting with the automatic photographic recording of meteor trails. He therefore asked Prentice for details of his Schmidt camera design (41). Since constructing a new instrument would inevitably take time, Lovell also enquired whether Prentice’s camera might once again be made available for use at Jodrell Bank.

The following year, Lovell and Prentice turned their attention to the potential of the newly developed photomultiplier technology for detecting faint meteors. Prentice opened discussions with EMI (42) regarding the construction of two image-converter systems (described in the correspondence as "image converter tubes", evidently an early form of image intensifier) to be used in conjunction with a military-surplus Kodak Aero Ektar lens (43). The intention was to install one instrument at Jodrell Bank and the second at Prentice's home in Suffolk (44).

To pursue the project, Prentice visited Dr J. D. McGee at the EMI Research Laboratories, Hayes, in December 1950 (45), taking the Aero Ektar lens with him so that it could be tested with the latest experimental image-converter tube then under development. As the technology remained subject to security restrictions, EMI required Admiralty approval before the equipment could be supplied, a matter that Lovell successfully resolved (46).

The instrument was subsequently completed and used successfully at Jodrell Bank to obtain photographic records of meteors in support of the radar observations.

Building on these encouraging results, Lovell commissioned the H. H. Wills Physical Laboratory at the University of Bristol—where he had completed his doctoral studies—to design and construct two specialised f/0.8 meniscus Super-Schmidt meteor cameras during the mid-1950s (47) (48). Their exceptionally fast optics and very wide fields of view made them ideally suited to meteor photography, and they became an integral part of the Jodrell Bank observing programme.

**Prentice worries that the days of visual meteor observing are drawing to a close**

In light of the rapid development of radar—which could operate day and night, regardless of the weather—Prentice became increasingly concerned about the future value of visual meteor observing. He was also aware of the growing importance of photographic techniques, which allowed meteor trajectories to be determined far more accurately than was possible by visual observation alone.

Prentice's doubts about the continuing role of visual observing first became apparent during 1949, and he shared these concerns with Lovell. The correspondence shows that Lovell repeatedly sought to reassure him. By October 1949, however, matters had come to a head, and Prentice outlined two principal concerns in a lengthy letter to Lovell: (49)

“1. The first is the breakdown in computing, and I don’t want to say much about this, as the computing has been Porter’s responsibility and we know that he has had a bad time since the end of the War.... The fact remains that if you cannot show that you are doing worth-while work your team will not hang together for ever, and the failure on the computing side has been a major factor in the break-up of the Section....

2. The second factor is a feeling that our work has become obsolete in the face of radio techniques.”

Prentice went on to explain that he had discussed the difficulties surrounding the computation programme with Porter and believed that a way forward might yet be found.

Lovell's replies, however, remained consistently supportive and encouraging. He emphasised the continuing importance of visual observations and the essential contribution that experienced visual observers could still make to the developing radar programme (31).

Nevertheless, the correspondence from the following June shows that Prentice's anxieties had not disappeared. He had previously agreed to prepare a memorandum outlining a future programme of coordinated visual and radio observations, but admitted that he had been unable to make progress (Figure 16): (50)

“I have yet to convince myself that there is going to be any more visual observing.... today I am still seeking for understanding whether my meteor work is finished and I am called to do other things.”

Lovell responded immediately with a lengthy typed reply, reassuring him: “I do not suffer from your doubts as to whether there will be any more visual work.” (51)

He went on to explain once again where he believed visual observers could continue to make an indispensable contribution before inviting Prentice to Jodrell Bank to discuss the matter with him and his colleagues: (51)

“I am sure that after such a visit you would need no further convincing that visual observing has to go on.”

Prentice accepted the invitation and visited Jodrell Bank on 17 July 1950. Much of their discussion centred on the practical difficulties of establishing a radar station in Suffolk, a proposal that Prentice now appeared to accept was unlikely to be realised.

A few days after the visit, Lovell wrote: (52) “As I see it the problem that you and your team were personally concerned [with] — that of the construction of the major showers and the minor streams — is a problem that remains with you.”

He continued by explaining how the visual observers could work even more closely with the Jodrell Bank researchers: (52)

“I do hope that the net result of all this is that you and your team will be inspired to carry forward this excellent work which it seems to me no one else in the world is likely to do. Clearly it can well be carried out by your team, and as far as we are concerned you will find that we shall give you every help...”

The exchange appears to have reassured Prentice, at least temporarily. The correspondence during the remainder of 1950 returned largely to practical matters: meteor observations, arrangements for BAA observers to visit Jodrell Bank (including George Alcock's visit that September), and plans for future collaborative projects, among them the image-converter camera system that Prentice was developing with EMI, described in the previous section.

**Prentice finally bows out**

The last letter I have found between Lovell and Prentice was written by Prentice on 18 December 1950. He had just returned from a successful Geminids observing expedition to Jodrell Bank and apologised for forgetting to return Lovell's deckchair, which he had borrowed for his night-time observations.

He ended the letter on an optimistic note: (53) “I hope that we shall be able during the coming year to carry out some further combined observations as I feel sure they have quite considerable value.”

Despite this, Prentice increasingly felt that the very techniques he had done so much to encourage had overtaken visual meteor observing, leaving little for the visual observer to contribute. At the same time, other responsibilities were placing increasing demands on his time. As he later reflected, he felt that he had been "called to do other things" (13).

A man of deep Christian faith, Prentice devoted considerable energy over the years to the Stowmarket Company of the Boys' Brigade and to his Congregational Church. The church building had been destroyed by bombing during the Second World War, and Prentice assumed a leading role in the construction of a new church in Stowmarket, which was opened in 1954. In the same year he was appointed Church Secretary, adding further to his already demanding workload.

Recognition of his astronomical achievements nevertheless continued. In 1953, the University of Manchester awarded him an honorary MSc in recognition of his pioneering work in visual meteor observation and his collaborative research with Jodrell Bank. In the same year, the Royal Astronomical Society awarded him the Hannah Jackson-Gwilt Medal "for his contribution to the study of meteors". (54)

After serving as Director of the BAA Meteor Section for 31 years, Prentice resigned in 1954. Although he announced that he would undertake no further observing, this proved not to be entirely the case. He returned to observe both the Perseids and Geminids in 1980, and the Perseids again in 1981 (55). He died just two months later.

Looking back on his collaboration with Lovell many years afterwards, Prentice wrote with evident regret:

“The war, which halted my work just as it was beginning to prosper, also accelerated the new techniques which were to replace it. But these new techniques meant that if I wished to continue seriously my research into meteors I must become a true scientist and devote my whole time to it. The opportunity, a fleeting one, did indeed occur. But I knew that my life had other purposes as well, which I could not forego, and in 1953 I gave up all my astronomical work, to my great and lasting sorrow”. (13)

In 1937 Prentice had been granted a Leverhulme Research Fellowship which enabled him to step away temporarily from his legal practice and, in April and May 1938, to travel to Madeira to investigate the radiant of the η Aquariid meteor shower (55) (13) (56). Whether this was indeed the opportunity he later had in mind cannot be known with certainty, but it illustrates that, at one stage, a professional career in meteor research had briefly appeared within reach.

Prentice was succeeded as Director of the Meteor Section by Harold Ridley, who inherited the Section at a difficult time when morale had declined. With strong encouragement from Lovell, Ridley refocused the programme on obtaining reliable observations of both shower and sporadic meteors while also embracing the emerging techniques of meteor photography and spectroscopy.

**Jodrell Bank after Prentice**

The research programme at Jodrell Bank also continued to evolve. Although meteor research remained an important part of its activities throughout the 1950s, increasing attention was devoted to the rapidly developing field of radio astronomy, in which Lovell and his colleagues were making internationally significant contributions.

As early as 1947, Lovell had constructed a 218-foot (66 m) transit radio telescope, consisting of a wire mesh reflector suspended from a circular framework of scaffolding. The instrument led to a series of important discoveries, including the first definitive detection of radio emission from the Andromeda Galaxy (M31), thereby establishing it as the first confirmed extragalactic radio source. It also detected radio emission from the remnant of Tycho's Supernova of 1572, before the source had been identified optically.

These successes marked Jodrell Bank's transition from a wartime radar station adapted for meteor research into one of the world's leading centres for radio astronomy. The next major step was the construction of the 250-foot (76 m) fully steerable radio telescope that now bears Lovell's name. Planning for the telescope began in 1949, and Lovell referred to the developing project in a letter to Prentice written in June 1950 (51).

The telescope became operational in 1957, just in time to track the launch rocket of *Sputnik 1*. Nearly seventy years later, it remains one of the world's leading research instruments and is the third-largest fully steerable radio telescope in operation.

Today, Jodrell Bank is home to the international headquarters of the Square Kilometre Array Observatory, which is constructing the world's largest radio telescope across sites in South Africa and Australia. In recognition of its outstanding scientific and historical significance, Jodrell Bank was designated a UNESCO World Heritage Site in 2019.

**Perspectives: a pioneering professional–amateur collaboration**

The story of the first meteor observations at Jodrell Bank is more than the story of a new scientific technique. It is also the story of a pioneering professional–amateur collaboration at a time when the boundaries between the two communities were becoming increasingly blurred.

Lovell brought to Jodrell Bank the expertise of a professional physicist, together with the new possibilities offered by wartime radar technology. Prentice brought decades of experience in meteor observation, an extensive network of skilled observers, and a deep understanding of the behaviour of meteors gained through years of patient visual observation. Neither could have achieved what they did alone. The success of the early Jodrell Bank meteor programme depended on the combination of their complementary skills.

The collaboration also demonstrates that amateur astronomers were not simply observers providing data for professional scientists. Prentice was an active scientific partner who helped shape the research programme, interpreted observations, contributed to publications and ensured that the new radar techniques remained connected to established meteor astronomy. The BAA observers who participated in the early Jodrell Bank observations played a similarly important role in validating and extending the discoveries.

The correspondence also reveals, however, some of the practical limitations of this collaboration. The BAA observers were dispersed across the country, most living in south-east England, making travel to Jodrell Bank both time-consuming and difficult. Professional and family commitments further limited the opportunities for prolonged visits. Prentice repeatedly expressed his own wish to spend more time at Jodrell, yet he was equally reluctant to remove his most experienced observers from their established observing stations during the major meteor showers, where their coordinated visual observations remained essential. These competing priorities inevitably constrained the extent of joint observing and may have limited the scientific opportunities that closer collaboration could have produced.

As Lovell's programme expanded, the Jodrell Bank team increasingly became able to operate independently. This gradual shift raises an interesting historical question: to what extent did the growing self-sufficiency of the professional team contribute to a perception among visual observers that their role was becoming less significant? The correspondence suggests that these concerns were felt particularly strongly by Prentice from about 1949 onwards. Whether this reflected the views of the Meteor Section observers more generally is less certain. Although several members of the Section continued to observe actively for many years, Prentice himself increasingly questioned the future of visual meteor astronomy and eventually withdrew from both active leadership of the Section and from observing.

Today, professional–amateur partnerships are recognised as an important component of modern astronomy, from variable star observations to exoplanet research. The collaboration between Lovell, Prentice and the BAA Meteor Section shows that this approach has a much longer history. The foundations of Jodrell Bank were built not only on innovative technology and professional expertise, but also on the enthusiasm, dedication and knowledge of amateur astronomers.

The first meteor echoes detected in that Cheshire field in December 1945 therefore represent more than the accidental discovery of a new astronomical phenomenon. They mark the beginning of a partnership that helped transform meteor science, contributed to the birth of radio astronomy, and demonstrated the enduring value of collaboration across the amateur–professional divide.

I encourage everyone to visit Jodrell Bank if they ever have the opportunity. Both Lovell and Prentice are celebrated in the exhibition in the First Light Pavilion, opened in 2022.

**Acknowledgements**

I am grateful to many people who have generously assisted with the research for this paper, in particular Bill Barton, James Dawson and Richard McKim, for their help in locating information and clarifying historical details. I also thank the librarians and archivists at the John Rylands Library, University of Manchester, for their invaluable assistance in accessing the Bernard Lovell Archive.

This paper is based on a talk presented by the author at the British Astronomical Association meeting held at the Institute of Physics, London, in June 2026.

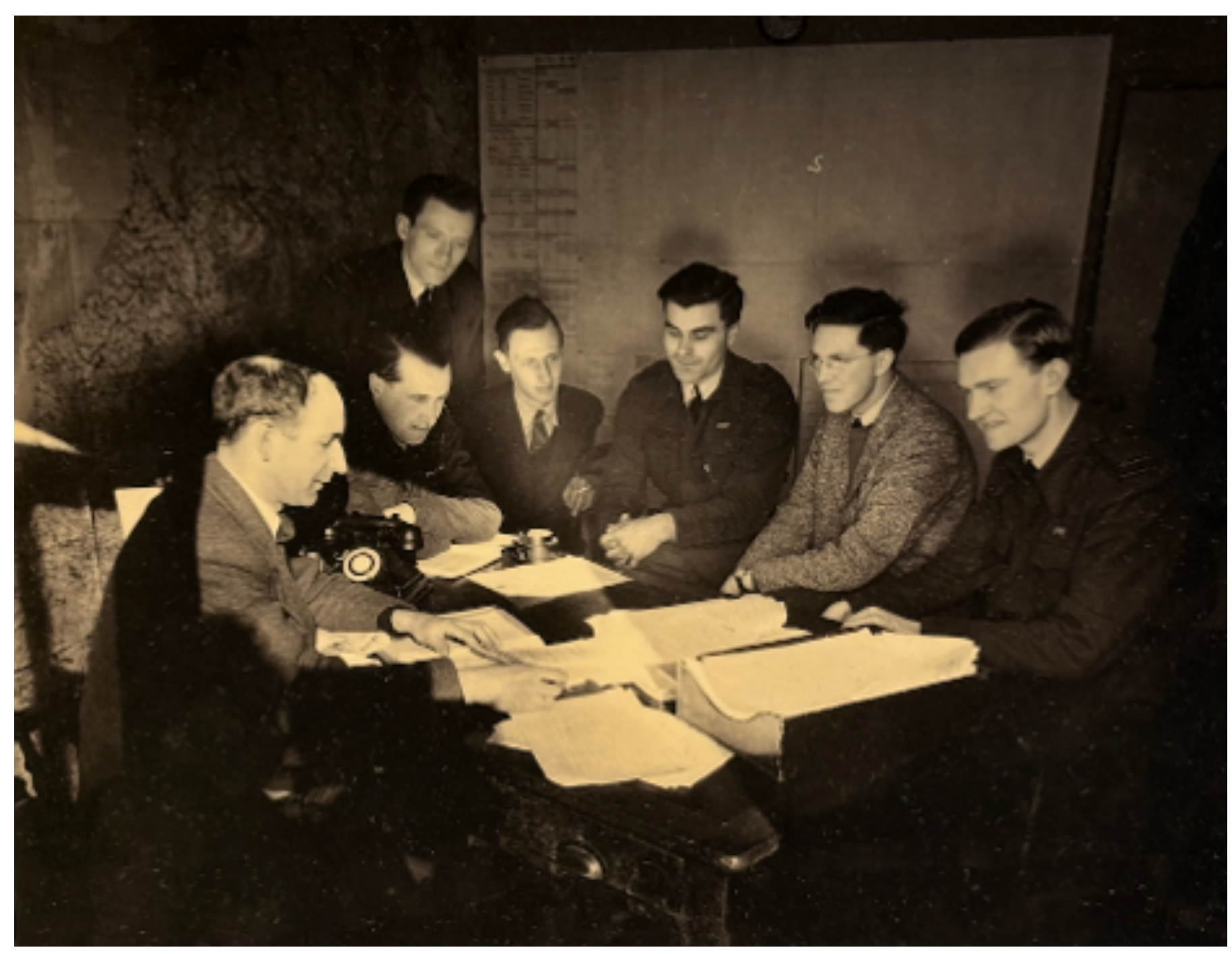

Figure 1: Lovell (left) and his H2S radar team at work, ca. 1942 (The University of Manchester)

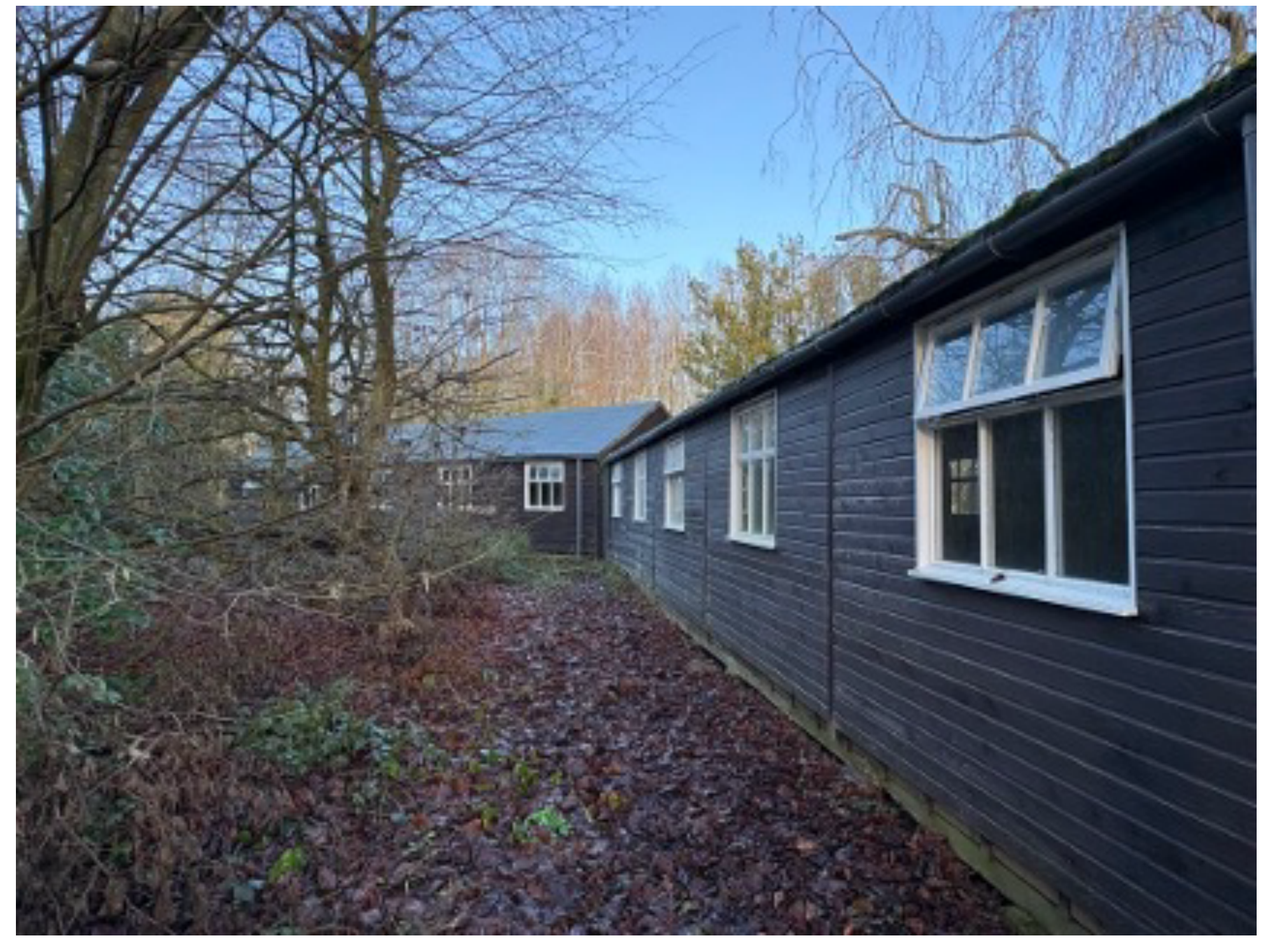

Figure 2: Gardeners' huts at Jodrell Bank in December 2025 (Jeremy Shears)

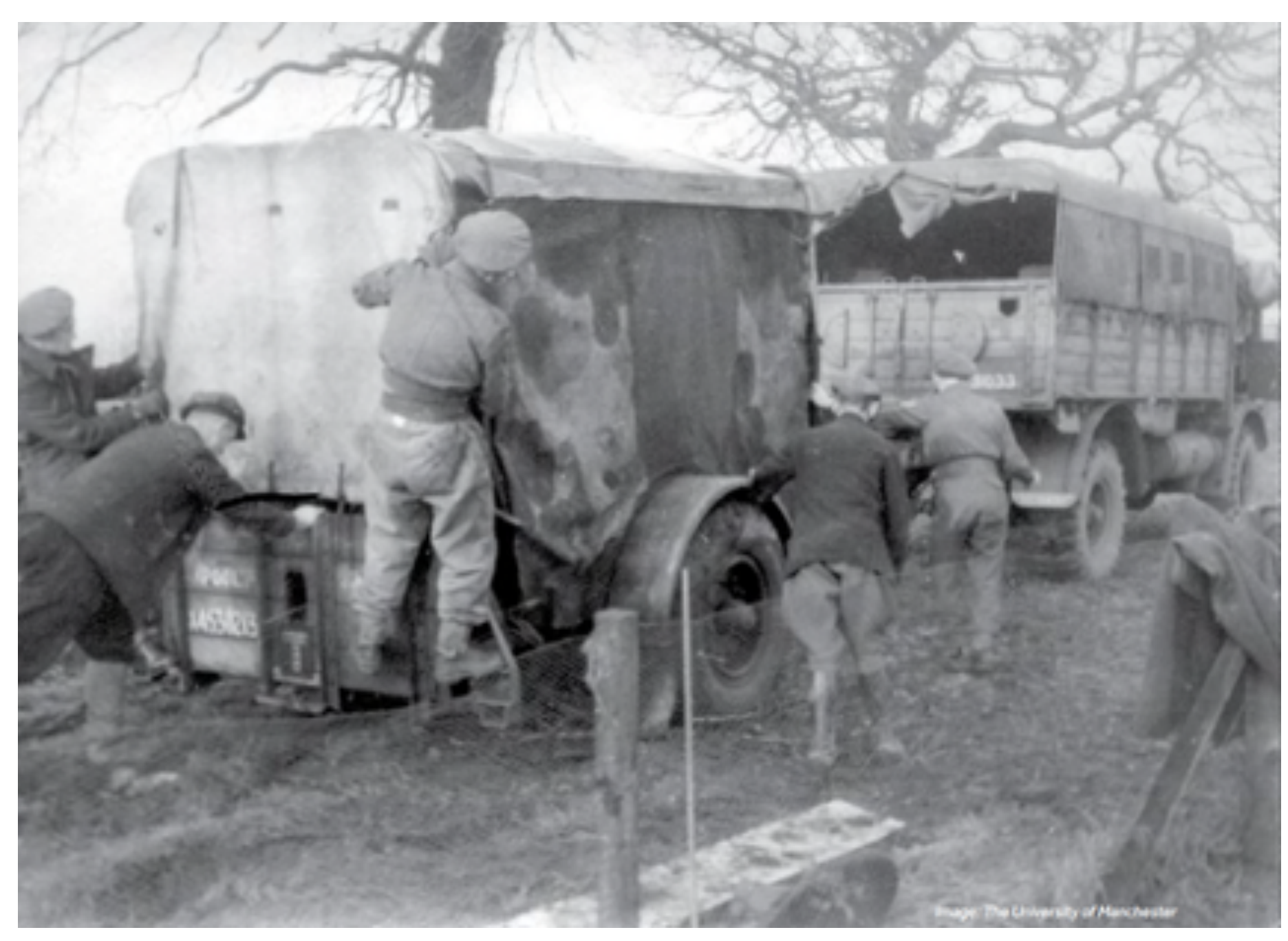

Figure 3: Army contingent moving radar equipment to Jodrell Bank, December 1945 (The University of Manchester)

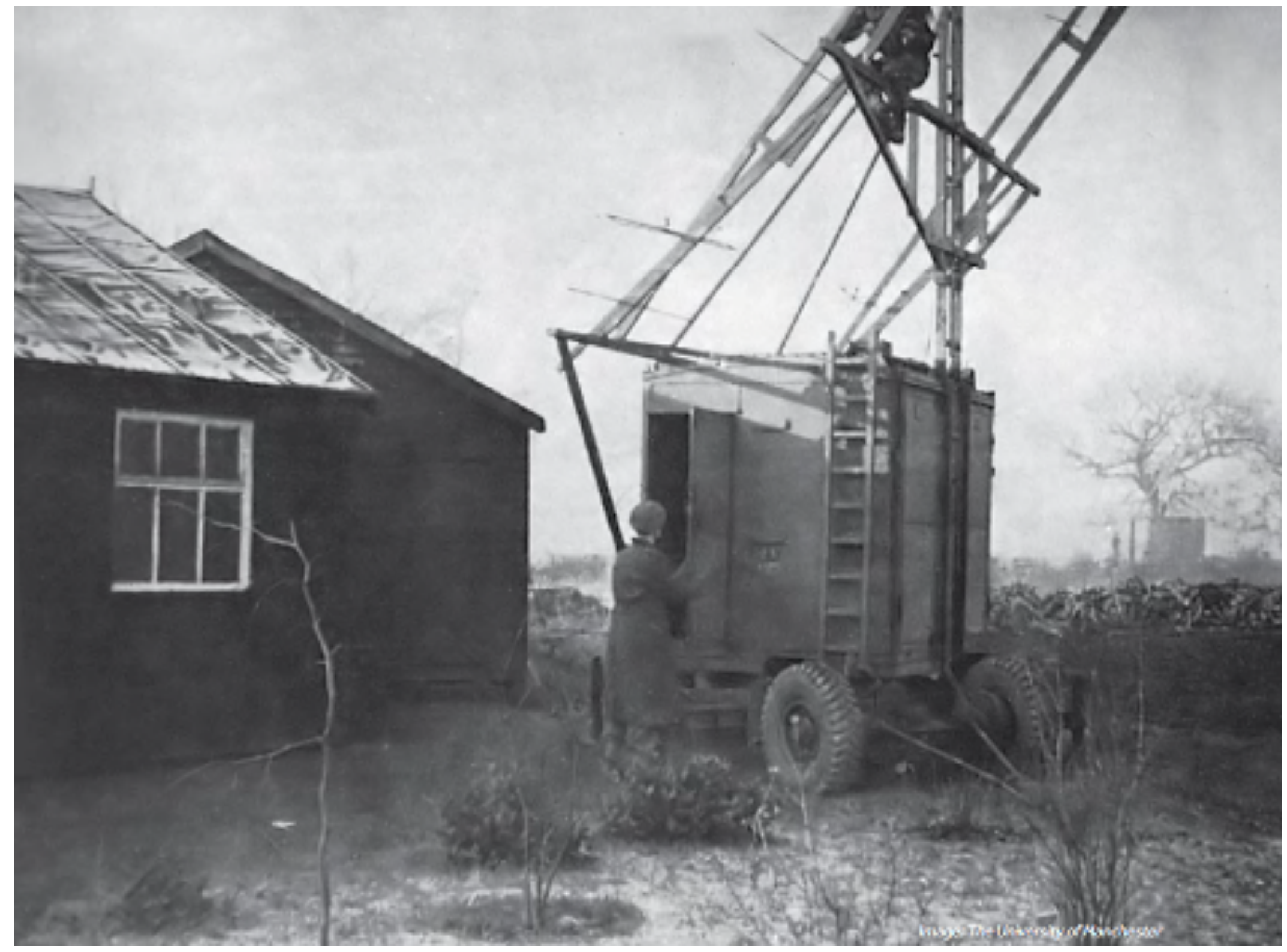

Figure 4: Installing radar aerials at Jodrell Bank, December 1945 (The University of Manchester)

Note that the two huts to the left are the same as in Figure 2.

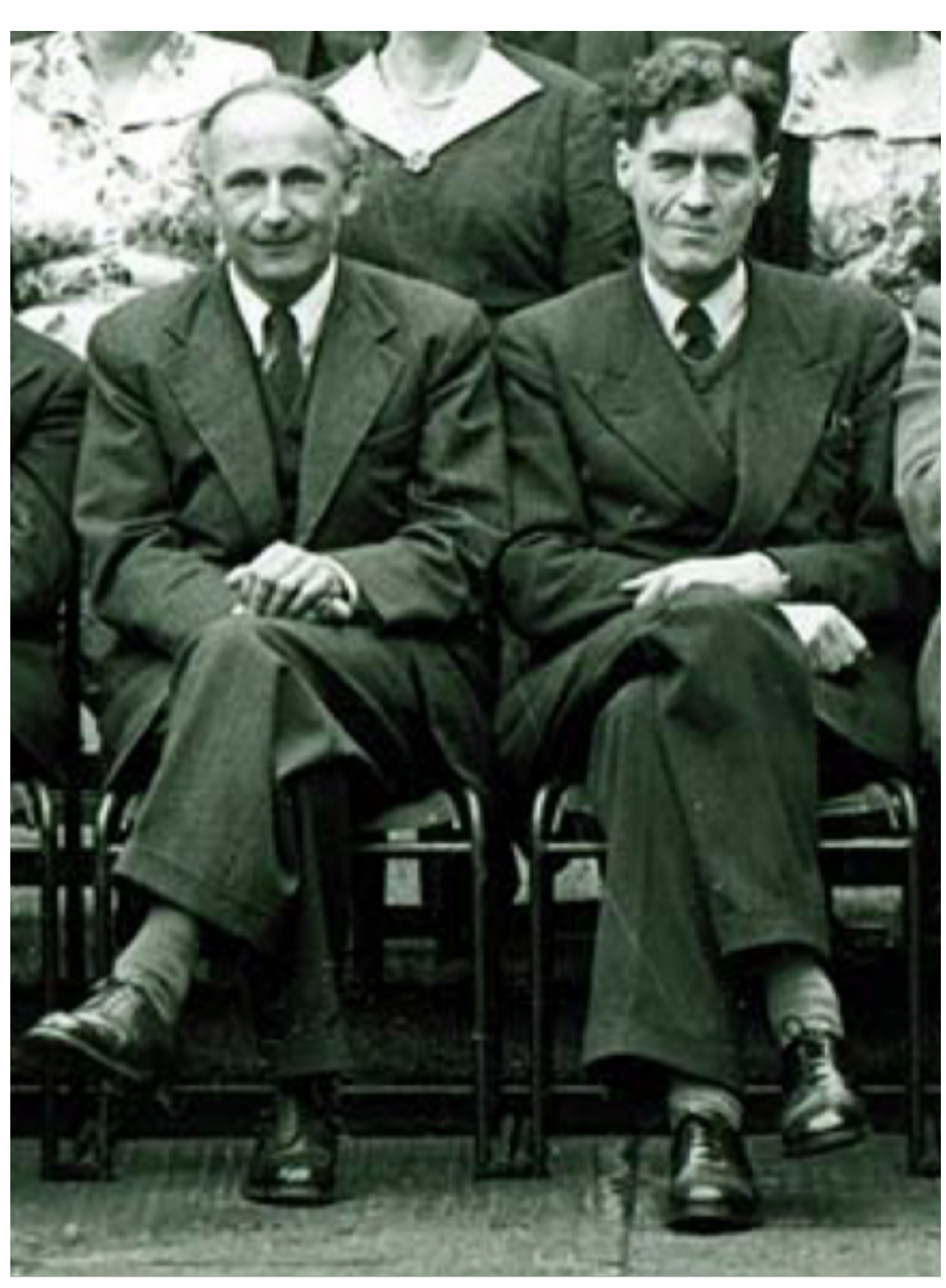

Figure 5: Lovell and Blackett in a group photograph of the University of Manchester's Physical Laboratories staff, 1951 (The University of Manchester)

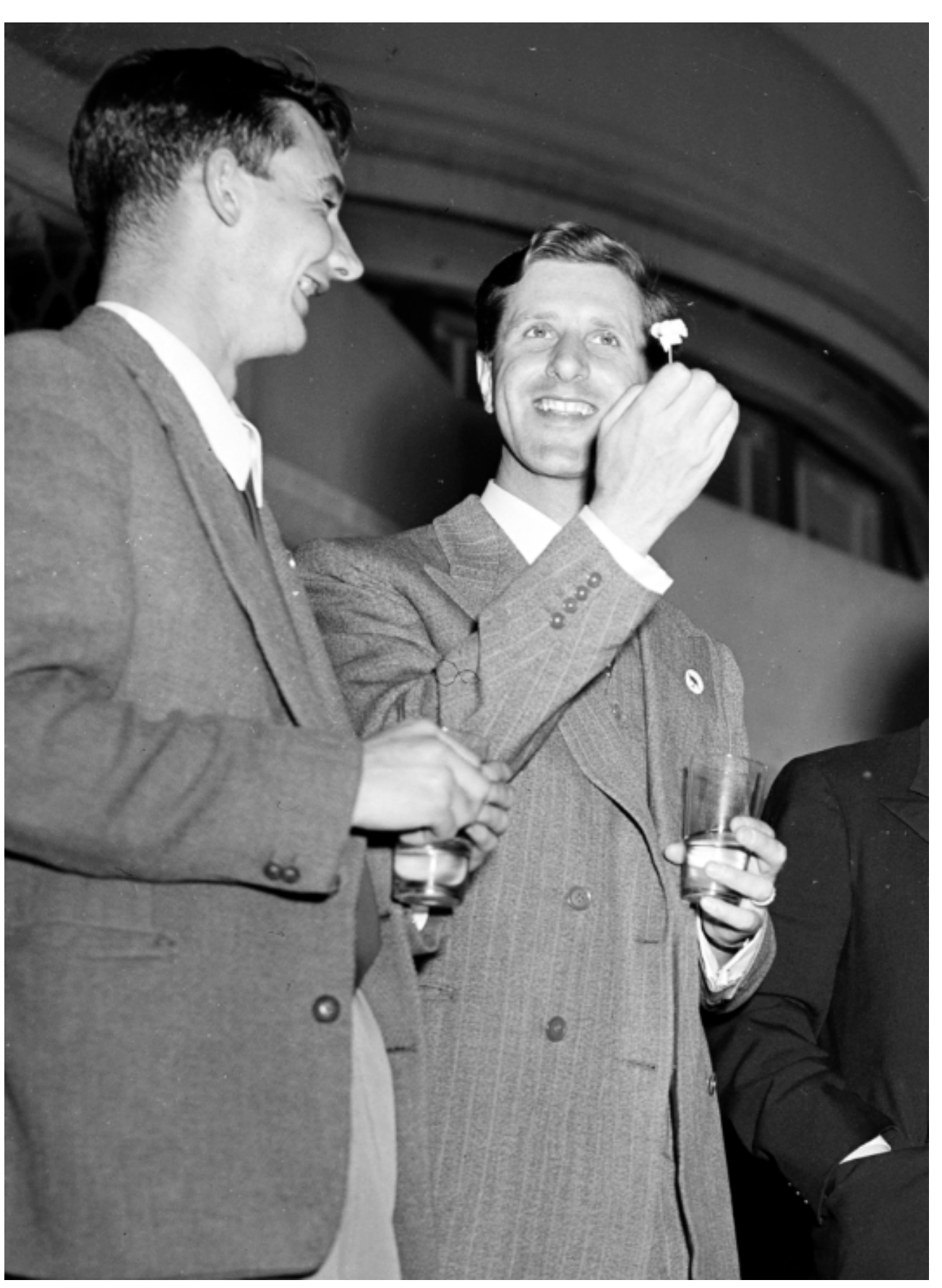

Figure 6: Nikolai Herlofson (right) at the 1952 URSI General Assembly in Sydney with radio astronomy pioneer, John Bolton (CSIRO Radio Astronomy Image Archive)

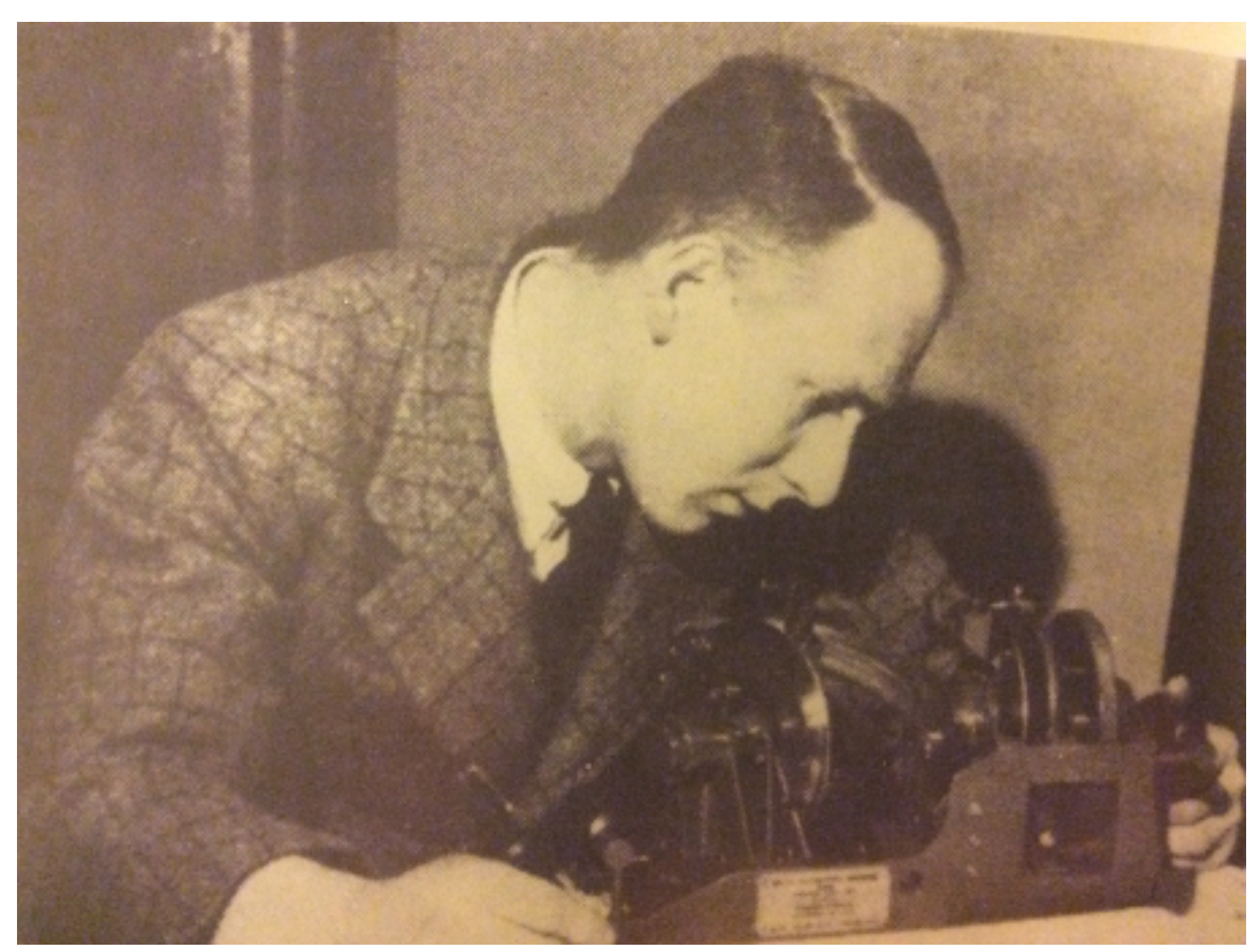

Figure 7: J.P. Manning Prentice in 1934

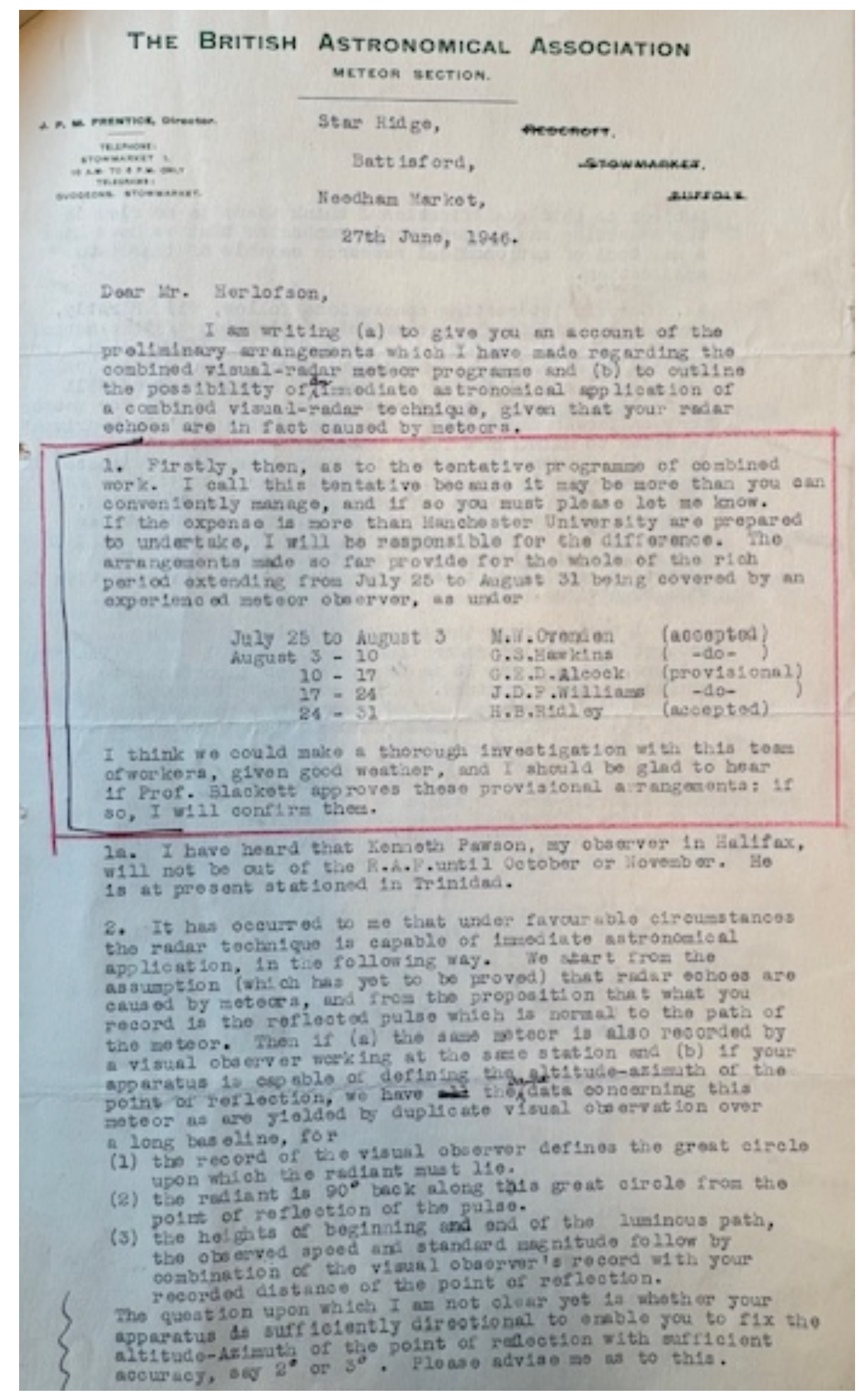

THE BRITISH ASTRONOMICAL ASSOCIATION
METEOR SECTION.

J. P. M. PRENTICE, Director.
TELEPHONE:
STOWMARKET

Star Ridge, ~~REDCROFT~~,
Battisford, ~~STOWMARKET~~,
Needham Market, ~~SUFFOLK~~

27th June, 1946.

Dear Mr. Herlofson,

I am writing (a) to give you an account of the preliminary arrangements which I have made regarding the combined visual-radar meteor programme and (b) to outline the possibility of immediate astronomical application of a combined visual-radar technique, given that your radar echoes are in fact caused by meteors.

1. Firstly, then, as to the tentative programme of combined work. I call this tentative because it may be more than you can conveniently manage, and if so you must please let me know. If the expense is more than Manchester University are prepared to undertake, I will be responsible for the difference. The arrangements made so far provide for the whole of the rich period extending from July 25 to August 31 being covered by an experienced meteor observer, as under

| | | |
|---|---|---|
| July 25 to August 3 | M.W.Cremden | (accepted) |
| August 3 - 10 | G.S.Hawkins | ( -do- ) |
| 10 - 17 | G.E.D.Alcock | (provisional) |
| 17 - 24 | J.D.P.Williams | ( -do- ) |
| 24 - 31 | H.B.Ridley | (accepted) |

I think we could make a thorough investigation with this team of workers, given good weather, and I should be glad to hear if Prof. Blackett approves these provisional arrangements: if so, I will confirm them.

1a. I have heard that Kenneth Pawson, my observer in Halifax, will not be out of the R.A.F. until October or November. He is at present stationed in Trinidad.

2. It has occurred to me that under favourable circumstances the radar technique is capable of immediate astronomical application, in the following way. We start from the assumption (which has yet to be proved) that radar echoes are caused by meteors, and from the proposition that what you record is the reflected pulse which is normal to the path of the meteor. Then if (a) the same meteor is also recorded by a visual observer working at the same station and (b) if your apparatus is capable of defining the altitude-azimuth of the point of reflection, we have the same data concerning this meteor as are yielded by duplicate visual observation over a long baseline, for

(1) the record of the visual observer defines the great circle upon which the radiant must lie.
(2) the radiant is 90° back along this great circle from the point of reflection of the pulse.
(3) the heights of beginning and end of the luminous path, the observed speed and standard magnitude follow by combination of the visual observer's record with your recorded distance of the point of reflection.

The question upon which I am not clear yet is whether your apparatus is sufficiently directional to enable you to fix the altitude-Azimuth of the point of reflection with sufficient accuracy, say 2° or 3°. Please advise me as to this.

Figure 8: Letter from Prentice to Herlofson detailing preliminary arrangements for observing the 1946 Perseids (Lovell Archive, University of Manchester)

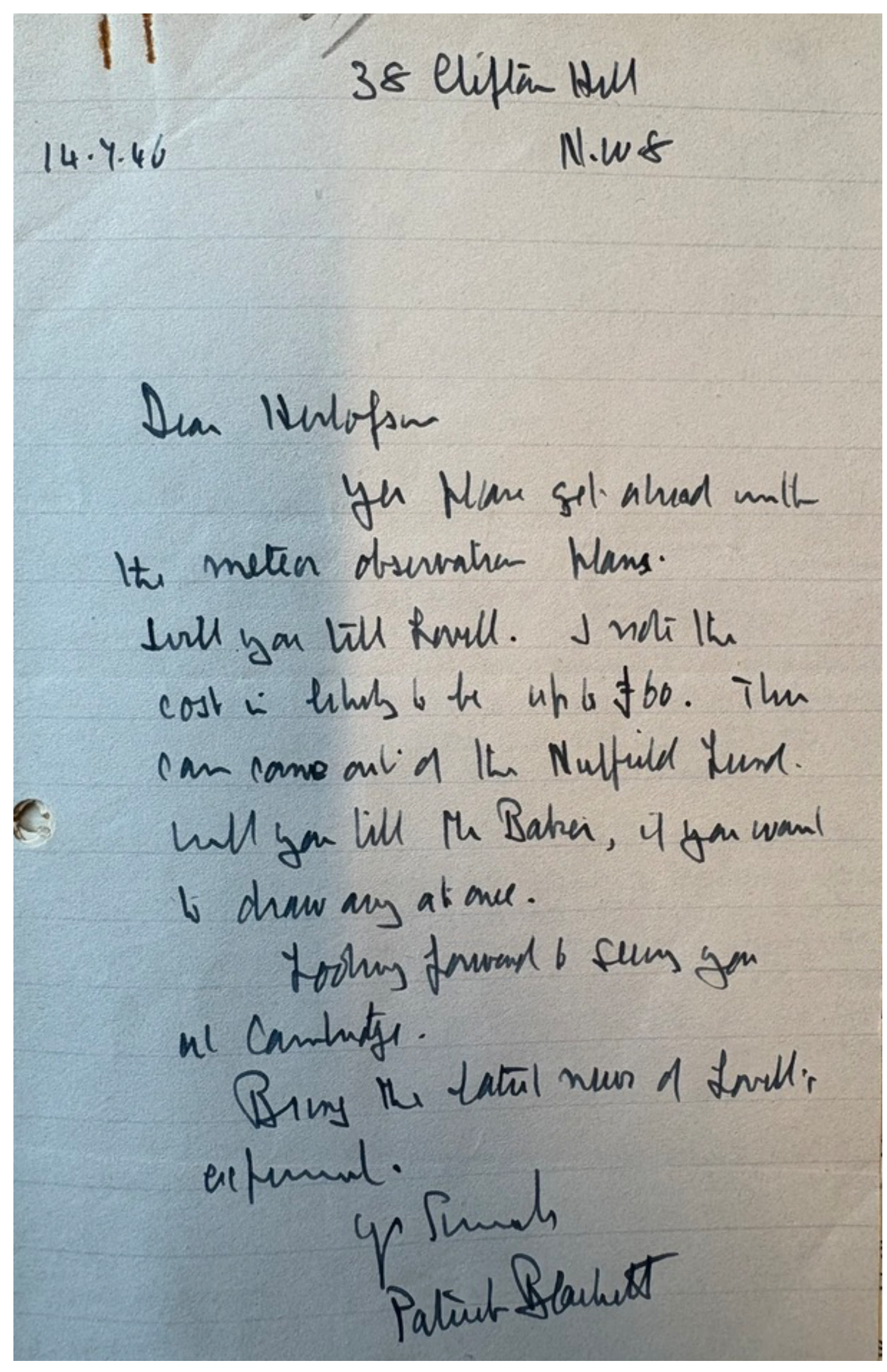

38 Clifton Hill
14.7.46 N.W.8

Dear Herlofson

Yes please get ahead with the meteor observation plans. Will you tell Lovell. I note the cost is likely to be up to £60. This can come out of the Nuffield Fund. Will you tell Mr Baker, if you want to draw any at once.

Looking forward to seeing you at Cambridge.

Bring the latest news of Lovell's experiment.

Yrs Sincerely
Patrick Blackett

Figure 9: Letter from Blackett to Herlofson approving funds for the BAA Meteor Section observers (Lovell Archive, University of Manchester)

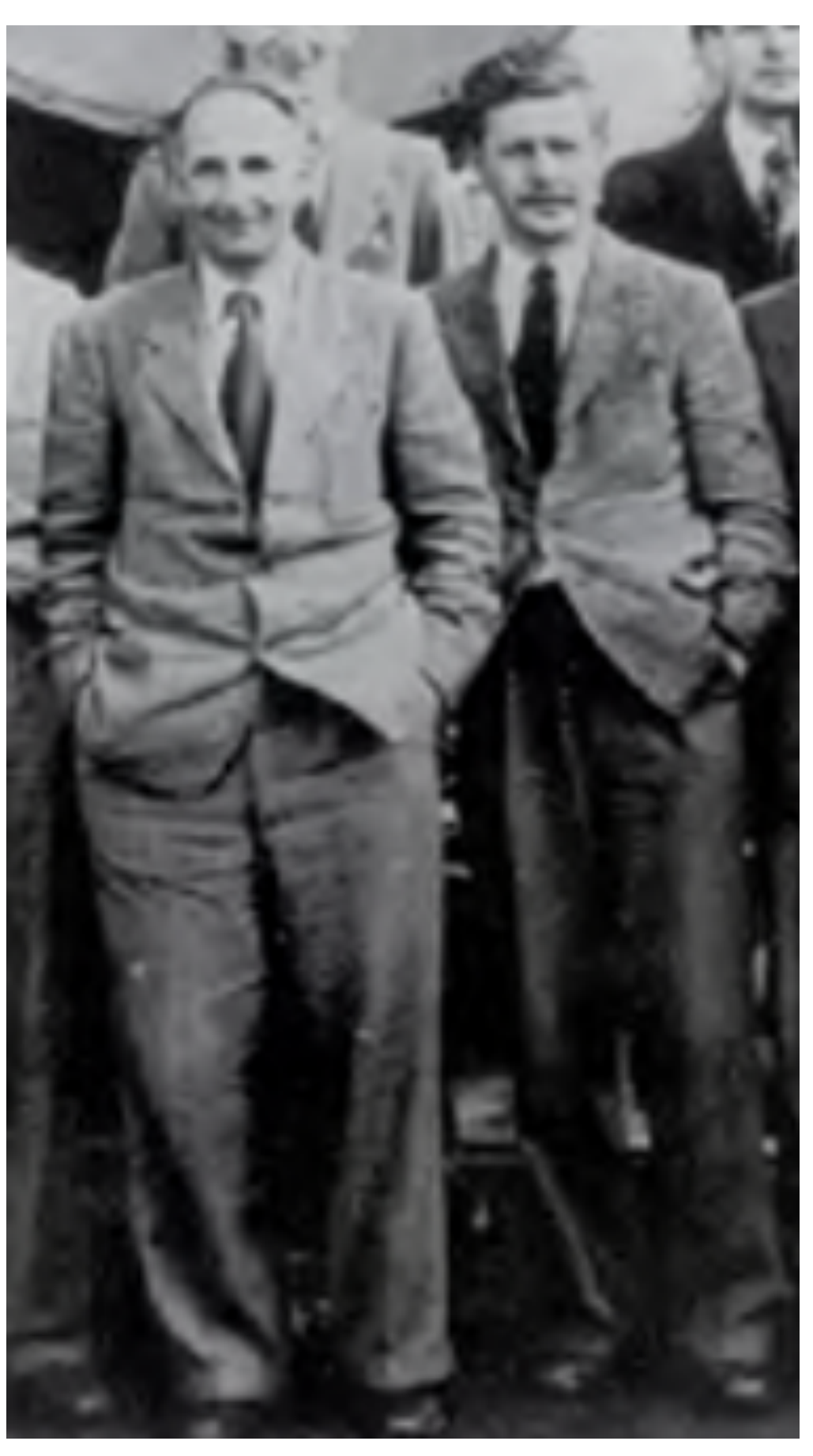

Figure 10: John Atherton Clegg, next to Lovell, in 1951 (The University of Manchester)

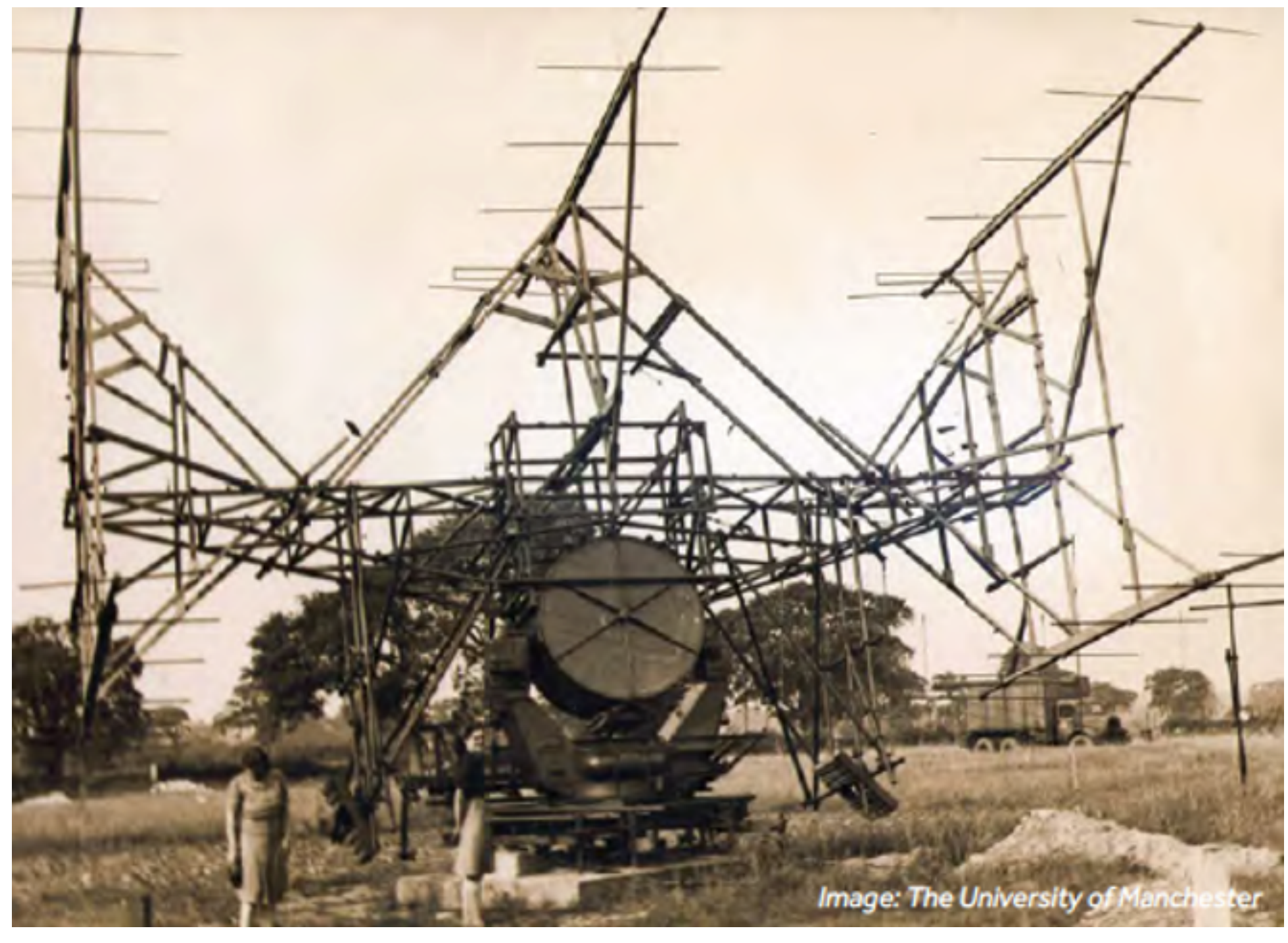

Figure 11: The Searchlight Aerial, ca. 1946. Note the ‘Park Royal’ truck in the background (The University of Manchester)

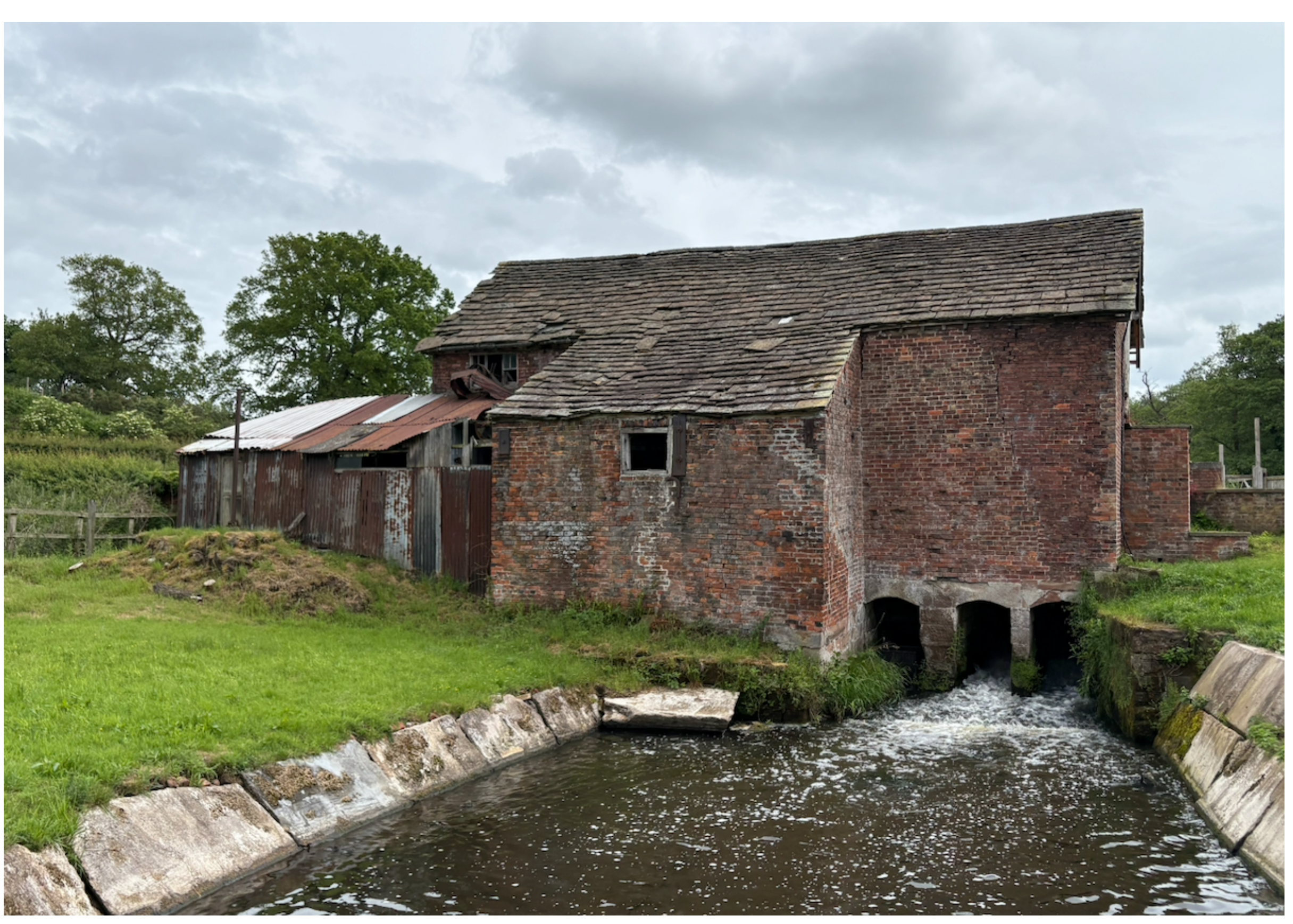

Figure 12: Bate mill at Peover Superior, near Jodrell Bank, June 2026 (Jeremy Shears)

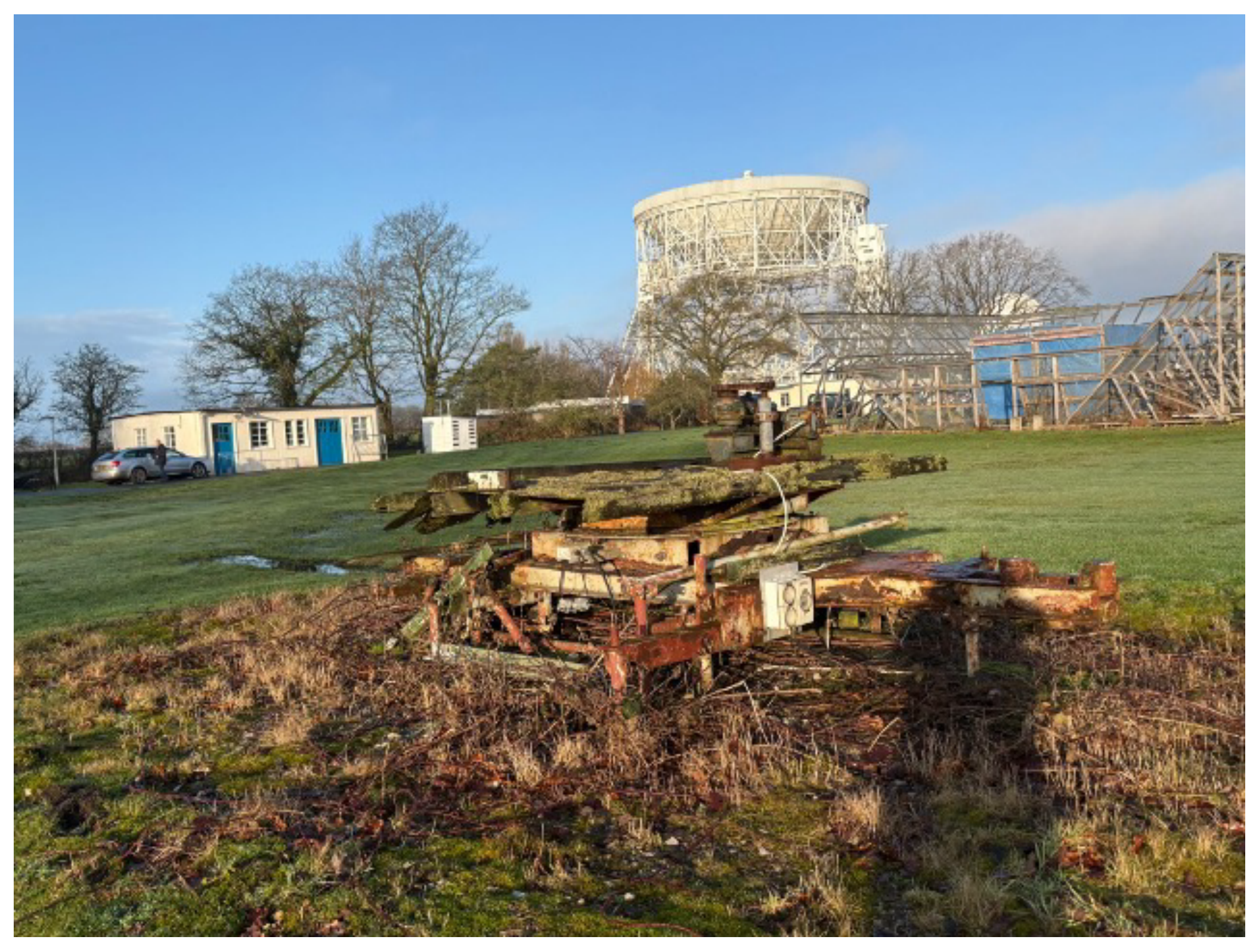

Figure 13: The Searchlight Aerial in 2025 (Jeremy Shears)

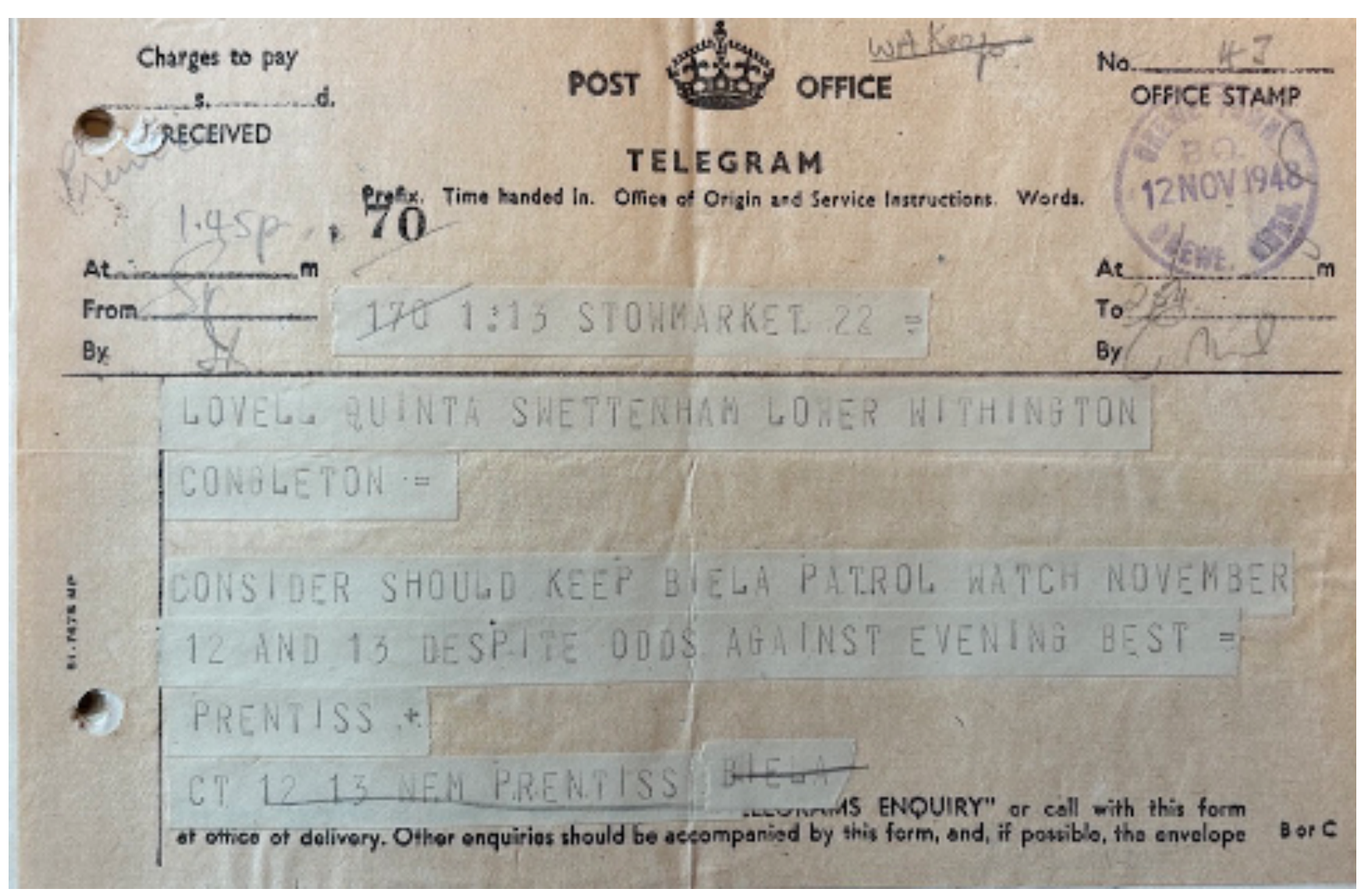

Charges to pay
s. d.
RECEIVED

POST OFFICE

TELEGRAM

No. 43
OFFICE STAMP
12 NOV 1948

Prefix. Time handed in. Office of Origin and Service Instructions. Words.
70

At m
From
By

170 1:13 STOWMARKET 22 =

At m
To
By

LOVELL QUINTA SWETTENHAM LOWER WITHINGTON
CONGLETON =

CONSIDER SHOULD KEEP BIELA PATROL WATCH NOVEMBER
12 AND 13 DESPITE ODDS AGAINST EVENING BEST =
PRENTISS +

CT 12 13 NEM PRENTISS BIELA

...S ENQUIRY" or call with this form at office of delivery. Other enquiries should be accompanied by this form, and, if possible, the envelope B or C

Figure 14: Telegram sent by Prentice in November 1948 alerting Lovell to the Bielids (The University of Manchester)

"Consider should keep Biela Patrol watch November 12 and 13 despite odds against. Evening best"

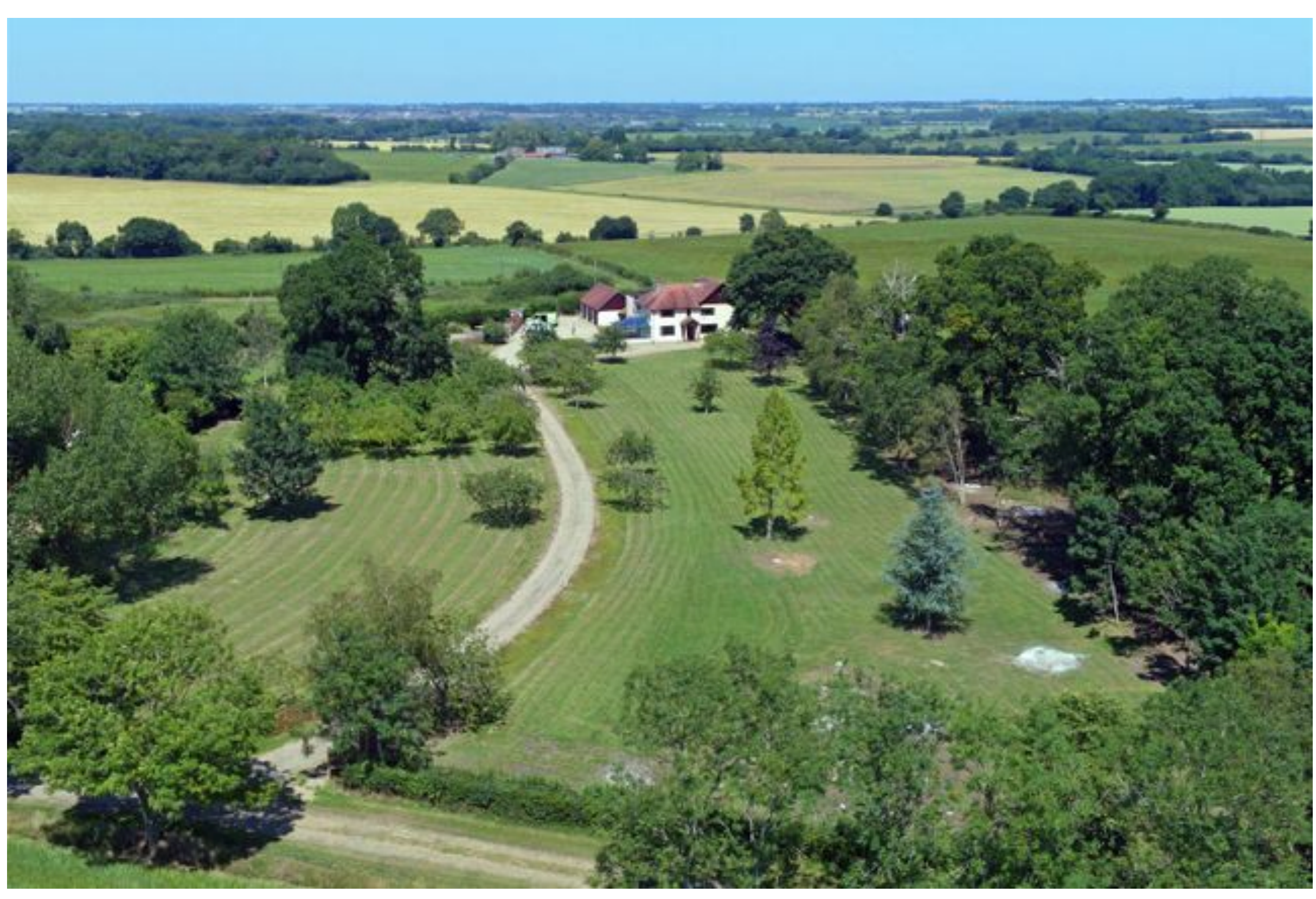

Figure 15: 'Star Ridge', Battisford, Suffolk – Prentice's former home

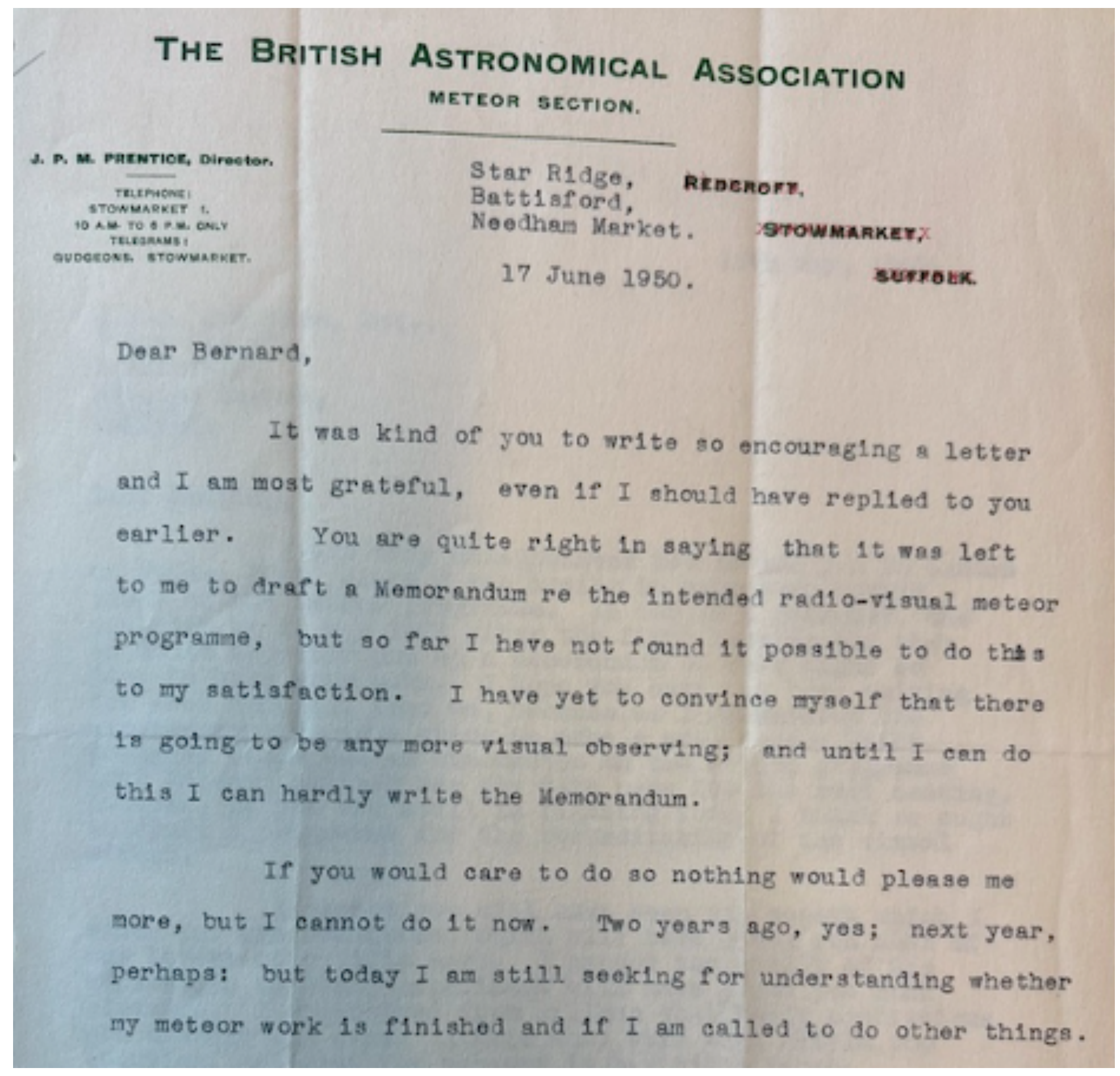
THE BRITISH ASTRONOMICAL ASSOCIATION
METEOR SECTION.

J. P. M. PRENTICE, Director.
TELEPHONE: STOWMARKET 1. 10 A.M. TO 6 P.M. ONLY
TELEGRAMS: GUDGEONS. STOWMARKET.

Star Ridge,
Battisford,
Needham Market.

~~REDCROFT, STOWMARKET, SUFFOLK.~~

17 June 1950.

Dear Bernard,

It was kind of you to write so encouraging a letter and I am most grateful, even if I should have replied to you earlier. You are quite right in saying that it was left to me to draft a Memorandum re the intended radio-visual meteor programme, but so far I have not found it possible to do this to my satisfaction. I have yet to convince myself that there is going to be any more visual observing; and until I can do this I can hardly write the Memorandum.

If you would care to do so nothing would please me more, but I cannot do it now. Two years ago, yes; next year, perhaps: but today I am still seeking for understanding whether my meteor work is finished and if I am called to do other things.

Figure 16: Letter from Prentice to Lovell, 17 June 1950, one of several in which he expresses doubts over the future of visual meteor work (The University of Manchester)